\documentclass[useAMS,usenatbib]{mn2e}
\usepackage{natbib,amssymb,amsmath,times,graphicx,url}
\usepackage{graphicx}
\usepackage{natbib}
\usepackage{subfigure}
\usepackage{placeins}
\usepackage{amsmath}
\usepackage{rotating}
\usepackage{url}
\usepackage{amssymb}
\input{psfig.sty}

\newcommand{\aips}{{$\cal AIPS\/$}}

\def\aap{A\&A}

\def\apjl{ApJ}
\def\apjs{ApJS}

\def\herschel{{\it Herschel}}
\def\spitzer{{\it Spitzer}}

\def\l1.4{$L_{\rm 1.4GHz}$}
\def\s1.4{$S_{\rm 1.4GHz}$}

\newcommand{\etal}{et~al.}
\newcommand{\msol}{\,\textrm{M}_{\odot}}                

\def\gs{\mathrel{\raise0.35ex\hbox{$\scriptstyle >$}\kern-0.6em
\lower0.40ex\hbox{{$\scriptstyle \sim$}}}}
\def\ls{\mathrel{\raise0.35ex\hbox{$\scriptstyle <$}\kern-0.6em
\lower0.40ex\hbox{{$\scriptstyle \sim$}}}}
\def\m@th{\mathsurround=0pt }
\def\eqalign#1{\null\,\vcenter{\openup1\jot \m@th
 \ialign{\strut\hfil$\displaystyle{##}$&$\displaystyle{{}##}$\hfil
 \crcr#1\crcr}}\,}
\title[The FIRRC in ALESS SMGs]{An ALMA survey of submillimetre galaxies in the Extended \textit{Chandra} Deep Field South: radio properties and the far-infrared/radio correlation}

\author[Thomson \etal]{A.\,P.\ Thomson,$^{\! 1}$ \thanks{E-mail: alasdair.thomson@durham.ac.uk}
  R.\,J.\ Ivison,$^{\! 2,3}$
  J.\,M.\ Simpson,$^{\! 1}$
  A.\,M.\ Swinbank,$^{\! 1}$
  Ian Smail,$^{\! 1}$
  V.\ Arumugam,$^{\! 2}$ \and
  D.\,M.\ Alexander,$^{\! 1}$
  A.\ Beelen,$^{\! 4}$
  W.\,N.\ Brandt,$^{\! 5,6}$
  I.\ Chandra,$^{\! 7}$
  H.\ Dannerbauer,$^{\! 8}$
  T.\,R.\ Greve,$^{\! 9}$ \and
  J.\,A.\ Hodge,$^{\! 10,11}$
  E.\ Ibar,$^{\! 12,13}$
  A.\ Karim,$^{\! 14}$
  E.\,J.\ Murphy,$^{\! 15}$
  E.\ Schinnerer,$^{\! 10}$
  S.\ Sirothia,$^{\! 7}$ \and
  F.\ Walter,$^{\! 10}$
  J.\,L.\ Wardlow,$^{\! 16}$
  \& P.\ van der Werf$^{17}$
  \vspace*{1mm}\\
  $^{1}$Institute for Computational Cosmology, Durham University, South Road, Durham DH1 3LE\\
  $^{2}$Institute for Astronomy, University of Edinburgh, Blackford Hill, Edinburgh EH9 3HJ\\
  $^{3}$European Southern Observatory, Karl-Schwarzschild-Str. 2, 85748 Garching bei M\"{u}nchen, Germany\\
  $^{4}$Institut d’Astrophysique Spatiale, b\^{a}t 121 - Universit\'{e} Paris-Sud, 91405 Orsay Cedex / France\\
  $^{5}$Department of Astronomy \& Astrophysics, 525 Davey Lab, The Pennsylvania State University, University Park, PA 16802, USA\\
  $^{6}$Institute for Gravitation and the Cosmos, The Pennsylvania State University, University Park, PA 16802, USA\\
  $^{7}$National Centre For Radio Astrophysics, Tata Institute Of Fundamental Research, Pune, India\\
  $^{8}$Universit\"{a}t Wien, Institut f\"{u}r Astrophysik, T\"{u}rkenschanzstra{\ss}e 17, 1180 Wien, Austria\\
  $^{9}$Department of Physics and Astronomy, University College London, Gower Street, London WC1E 6BT, UK\\
  $^{10}$Max-Planck Institute for Astronomy, K\"{o}nigstuhl 17, D-69117 Heidelberg, Germany\\
  $^{11}$National Radio Astronomy Observatory, 520 Edgemont Road, Charlottesville, VA 22901 (USA)\\
  $^{12}$Instituto de F\'isica y Astronom\'ia. Universidad de Valpara\'iso. Avda. Gran Breta\~na 1111. Valpara\'iso. Chile\\
  $^{13}$Instituto de Astrof\'isica. Facultad de F\'isica. Pontificia Universidad Cat\'olica de Chile. Casilla 306, Santiago 22, Chile\\
  $^{14}$Argelander-Institute for Astronomy, Bonn University, Auf dem H\"{u}gel 71, D-53121 Bonn, Germany\\
  $^{15}$Infrared Processing and Analysis Center, California Institute of Technology, MC 220-6, Pasadena CA, 91125, USA\\
  $^{16}$Dark Cosmology Centre, Niels Bohr Institute, University of Copenhagen, Denmark\\
  $^{17}$Leiden Observatory, Leiden University, P.O. Box 9513, 2300 RA Leiden, The Netherlands
}

\date{Accepted 2014 04 29}
\pagerange{\pageref{firstpage}--\pageref{lastpage}} \pubyear{2014}

\begin{document}
\label{firstpage}
\bibliographystyle{mnras}
\maketitle

\begin{abstract}
We present a study of the radio properties of 870\,$\mu$m-selected submillimetre galaxies (SMGs), observed at high resolution with ALMA in the Extended Chandra Deep Field South. From our initial sample of 76 ALMA SMGs, we detect 52 SMGs at $>3\sigma$ significance in VLA 1400\,MHz imaging, of which 35 are also detected at $>3\sigma$ in new 610 MHz GMRT imaging. Within this sample of radio-detected SMGs, we measure a median radio spectral index $\alpha_{610}^{1400} = -0.79 \pm 0.06$, (with inter-quartile range $\alpha=[-1.16,-0.56]$) and investigate the far-infrared/radio correlation via the parameter $q_{\rm IR}$ , the logarithmic ratio of the rest-frame 8--1000\,$\mu$m flux and monochromatic radio flux. Our median $q_{\rm IR} = 2.56 \pm 0.05$ (inter-quartile range $q_{\rm IR}=[2.42,2.78]$) is higher than that typically seen in single-dish 870\,$\mu$m-selected sources ($q_{\rm IR} \sim 2.4$), which may reflect the fact that our ALMA-based study is not biased to radio-bright counterparts, as previous samples were. Finally, we search for evidence that $q_{\rm IR}$ and $\alpha$ evolve with age in a co-dependent manner, as predicted by starburst models: the data populate the predicted region of parameter space, with the stellar mass tending to increase along tracks of $q_{\rm IR}$ versus $\alpha$ in the direction expected, providing the first observational evidence in support of these models.
\end{abstract}

\begin{keywords}
galaxies: starburst, galaxies: evolution, galaxies: high-redshift
\end{keywords}

\section{Introduction}\label{sect:intro}

Since their discovery in the late 1990s, it has become apparent that submillimetre-selected galaxies \citep[SMGs:][]{blain02,casey13} are massive, high-redshift ($z\sim2$), far-infrared luminous ($L_{\rm IR} \gtrsim 10^{12}$\,L$_{\odot}$) galaxies, which host prodigious dust-enshrouded starbursts with star-formation rates (SFR) as high as $\sim 10^3$\,M$_{\odot}$\,yr$^{-1}$. SMGs are a thousand times more numerous at $z\sim 2$ than are similarly (ultra-)luminous infrared galaxies at $z\sim 0$ \citep[ULIRGs:][]{sanders96}, and are believed to account for a significant fraction ($\gtrsim 20$\%) of the star formation rate density at $z\sim 2$ \citep[e.g.][]{hughes98, casey13, swinbank13}. Nevertheless, detailed understanding of SMGs and the role they play in the build-up of stellar mass in the Universe has proven challenging, due largely to a combination of the poor angular resolution (and the resulting high confusion noise) of the single dish maps in which SMGs are first detected, and to the faintness of SMGs in bands other than the submillimetre, both of which make robust and unambiguous identification of counterparts at other wavelengths difficult. 

Radio imaging has provided one means of identifying samples of SMGs at high-angular resolution \citep[e.g.][]{ivison98, smail00}, via the assumption that they obey the observed correlation between far-infrared (far-IR) and radio luminosities seen in star-forming galaxies \citep[e.g.][]{dejong85, appleton04}. This far-IR/radio correlation (FIRRC) is believed to arise due to a balance between two key processes in the lives of massive ($M>8$\,$\msol$) stars: emission in the far-infrared is produced when their UV/optical photons are absorbed and reprocessed by dust, while the radio luminosity is powered predominantly via synchrotron emission of cosmic ray electrons and positrons, caught in the magnetic fields of supernova remnants, formed when these same massive stars die \citep{voelk89,lacki10}. Balance is achieved between these two emission processes when the duration of the starburst exceeds the typical lifetimes of these massive stars ($>10$\,Myr).

The tightness of the FIRRC has made it a valuable tool in the study of galaxy evolution, having been used: (i) to estimate the SFR of dusty nuclear starbursts \citep{condon92}; (ii) as an age-estimator for starbursting galaxies, based on whether they show excess infrared or radio emission \citep{bressan02}; (iii) as a means of searching for radio-excess AGN \citep{donley05, delmoro13}; and (iv) as a very crude redshift estimator \citep[SMGs: ][]{carilli99}.

In recent years, efforts have been made to push studies of the FIRRC to higher redshifts, and in particular, to test whether, it evolves with redshift \citep[e.g.][]{garrett02, boyle07, ibar08, seymour09, sargent10, bourne11, barger12}, as may be expected given the dependence of the inverse Compton cooling rate on the temperature of the Cosmic Microwave Background \citep{carilli99, schleicher13}. Notably, \citet{ivison10a} studied the evolution with redshift of the logarithmic ratio 

\begin{equation}
q_{\rm IR}= \log_{10}\biggl[\frac{S_{\rm IR}}{3.75\times 10^{12}{\rm \,W\,m}^{-2}}\times \frac{{\rm W\,m}^{-2}{\rm \,Hz}^{-1}}{S_{\rm 1.4\,GHz}}\biggr]
\end{equation}

\noindent where $S_{\rm IR}$ is the rest-frame 8-1000$\mu$m flux and $S_{\rm 1.4 GHz}$ is the rest-frame 1.4-GHz flux density, finding a modest evolution $q_{\rm IR} \propto (1+z)^{\gamma}$ with $\gamma =-0.15 \pm 0.03$, \citep[later revised to $\gamma =-0.26 \pm 0.07$ as \herschel\ data became available;][]{ivison10b}. 

These studies have, however, suffered from several limitations. Firstly, as noted above, the poor angular resolution  ($\sim 10$--$20''$) of the single-dish submillimetre maps in which SMGs are first detected makes it difficult to locate the starburst which powers their prodigious far-IR emission, and has the potential to blend several distinct sources, leading to over-estimation of the far-infrared luminosities of any identified counterparts \citep[e.g.][]{hodge13}. Secondly, the use of radio IDs to locate counterparts to submillimetre sources introduces a strong bias towards those sources which happen to be radio-luminous, and additionally introduces a redshift bias due to the strong $K$-correction experienced in the radio bands. Thirdly, in order to determine the radio luminosities of galaxies, it is necessary to $K$-correct the observed flux densities assuming $S_{\nu} \propto \nu^{\alpha}$, where the spectral index $\alpha$\ is typically unconstrained. While $\alpha=-0.7$ or $-0.8$ is commonly assumed \citep[based on the measurements of][]{ibar10}, steep spectra ($\alpha < -1.0$) are sometimes seen in early-stage mergers, and flat or inverted spectra ($\alpha > 0$) can be associated with the large free-free optical depths of fully coalesced mergers \citep[e.g.][]{murphy13}, or with dominant, strongly synchrotron self-absorbed AGN \citep{kaiser06}. Hence the assumption of a single radio spectral index for all SMGs is a simplification which may bias the interpretation of the FIRRC. 

In this paper, we circumvent these limitations by bringing together high-angular resolution 870\,$\mu$m observations from the Atacama Large Millimeter Array (ALMA), 1.4\,GHz observations from the Karl G. Jansky Very Large Array (VLA) and new 610\,MHz Giant Metre-Wave Radio Telescope (GMRT) observations covering the Extended \textit{Chandra} Deep Field South (ECDFS). The ALMA observations allow us to obtain precise positions for a sample of SMGs, allowing us to unambiguously identify their radio counterparts without reliance on statistical associations to mid-IR/radio counterparts. We then combine the two radio maps to determine individual spectral indices for these SMGs, allowing us to individually $K$-correct their observed radio flux densities (without having to assume an average spectral index), and thus measure $q_{\rm IR}$ for a sample of SMGs free from the biases inherent in using prior radio-identified counterparts. Throughout this paper, we use a cosmology with $H_{\rm 0}=71$\,km\,s$^{-1}$\,Mpc$^{-1}$, $\Omega_{\rm m}=0.27$, and $\Omega_{\Lambda}=0.73$, which gives an angular scale of 8.5\,kpc\,arcsec$^{-1}$ at $z=2$.

\section{Observations and sample selection}
\subsection{ALMA 870-$\mu$m observations}

These observations were presented in \citet{hodge13}, in which detailed descriptions of the data reduction and imaging are given. Briefly, we observed 122 of the 126 known submillimetre sources from the LABOCA ECDFS (LESS) survey of \citet{weiss09} with ALMA's Band 7 receivers (870-$\mu$m) in Cycle 0, between 2011 October 18 -- November 3. With $\sim15$ antennas available for each track, and using {\sc clean} with natural weighting, we achieve a typical synthesized beam of $\sim 1.6\times1.2''$, and with on-source integration times of $\sim120$\,s per source, reach a typical RMS sensitivity of $\sigma \sim 0.4$\,mJy\,beam$^{-1}$, some 3$\times$ deeper and with a beam area 200$\times$ smaller than the original single-dish LESS observations. 

Of these 122 ALMA maps, 88 have a primary beam corrected RMS $< 0.6$\,mJy and a beam axial ratio $<2.0$. We extract 99 SMGs at S/N $>3.5$ from 69 of these 88 maps -- the ALESS {\sc main} sample -- while 17 of the maps ($20\%$) contain no ALMA counterpart, most likely due to multiplicity effects \citep[see][]{karim13, hodge13}.

\subsection{GMRT 610-MHz observations}

GMRT observations of the ECDFS were obtained between 2009 October and 2010 February. During eleven $\sim$6.5\,hr sessions we observed six positions, placed equidistant on a circle of radius 0.2\,deg, centred on 03h\,32m\,28.0s $-27^\circ 48' 30''$ (J\,2000). The total bandwidth was 32\,MHz, split into 512 channels, centred at 610\,MHz, recording RR and LL polarisations with a time resolution of 2\,s.

Data reduction was accomplished using \aips. 3C\,48 or 3C\,147 were used to calibrate the bandpass and set the flux-density scale; 0240$-$231 was used to track the complex gain. Data corrupted by radio-frequency interference (RFI) were excised using channel- and time-based flagging. The final mosaiced image, made using the {\sc Wproject} algorithm after several rounds of self-calibration in phase, and one round of self-calibration in amplitude, reaches an RMS sensitivity of $45$\,$\mu$Jy\,beam$^{-1}$ over $70'\times 70'$, encompassing the positions of all 99 ALESS {\sc main} SMGs. Multiple facets were created using the wide-field Clark {\sc clean} algorithm with Briggs weighting ({\sc rmode} = `norm', {\sc robust} = 0). The synthesised beam measured $4.4'' \times 3.0''$ at position angle, $24.3^\circ$.

\subsection{VLA 1.4-GHz observations}\label{sect:vla}
We rely on data from the VLA 1.4-GHz Survey of the ECDFS DR2 \citep{miller13}, a re-reduction of the same 250\,h of A-configuration data taken from 2007 June -- September as were used by \citet{biggs11} to produce the first deep radio catalogue of ECDFS. This updated map covers an area of a third of a square degree, and reaches an RMS sensitivity at the centre of the field of 6\,$\mu$Jy\,beam$^{-1}$. 93 of the 99 ALESS {\sc main} SMGs lie within the VLA coverage. The angular resolution, $2.8'' \times 1.6''$,  closely matches that of the ALMA observations, and we use this new map to directly identify radio counterparts to these SMGs. 

In order to account for the effects of bandwidth smearing (BWS) -- which tend to decrease the peak flux density and elongate sources radially towards the phase centre of each pointing in wide-bandwidth interferometer images \citep{bridle89} -- we use information contained within the DR2 image header and follow the technique of Arumugam et al.\ (2014, \textit{in prep}) to create a map of BWS as a function of position, from which we measure typical flux corrections of 10--15\,\%.

\subsection{Optical--far-IR data and photometric redshifts}\label{sect:ir}

In our analysis we exploit previously-measured photometric redshifts and far-IR luminosities. Here, we briefly summarise these data. 

\citet{simpson13} derived photometric redshifts ($z_{\rm phot}$) for the ALESS SMGs utilising panchromatic imaging from the optical to mid-infrared. This comprises $U$ to $K$-band imaging from the MUltiwavelength Survey by Yale-Chile \citep[MUSYC:][]{gawiser06}, $J$ \& $K$ imaging from ESO-VLT/HAWK-I (Zibetti \etal, \textit{in prep}) and the Taiwan ECDFS NIR Survey \citep[TENIS:][]{hsieh12}, and images from the \spitzer\ IRAC/MUSYC Public Legacy in ECDFS \citep[SIMPLE:][]{damen11} at 3.6, 4.5, 5.8 and 8.0\,$\mu$m. Using the {\sc hyperz} code of \citet{bolzonella00}, they estimate redshifts for 78/99 ALESS {\sc main} SMGs, with a median uncertainty $\Delta z = \pm 0.21$. 17 of these SMGs lie at $z>3$.  The remaining 21 {\sc main} SMGs have insufficient optical/near-IR photometry to measure reliable redshifts, but are statistically expected to lie at $z\sim2.5$--4.0. Full details of this redshift fitting procedure and of the $z_{\rm phot}$ distribution and comparison with available spectroscopic redshifts are presented in \citet{simpson13}.

To derive the far-infrared (FIR) properties of the ALESS SMGs, \citet{swinbank13} fit the deblended 250--500\,$\mu$m \herschel\ SPIRE photometry \citep{pilbratt10, griffin10} with a suite of dust templates at the appropriate $z_{\rm phot}$ in the range $T_{\rm d}$\,=\,19--60\,K. The resulting infrared luminosities span $L_{\rm IR}=0.2$--$15\times10^{12}$L$_{\odot}$, with a median $\langle L_{\rm IR} \rangle =3\times10^{12}$L$_{\odot}$. Full details of the deblending process, and of the derived far-IR properties of each ALESS SMG are presented in \citet{swinbank13}.

\subsection{Final sample}
For the remainder of this work, we restrict our analysis to the 76 ALMA 870\,$\mu$m-detected SMGs with measured $z_{\rm phot}$ \citep{simpson13} and deblended far-IR luminosities, $L_{\rm IR}$ \citep{swinbank13}, which lie within the area covered by the VLA 1.4\,GHz map. We refer to this as the ``full sample'', and identify VLA-detected and GMRT-detected sub-samples as appropriate throughout the analysis. This sample selection is summarized in Table\,1.

\begin{figure*}
\centering
\includegraphics[width=\textwidth]{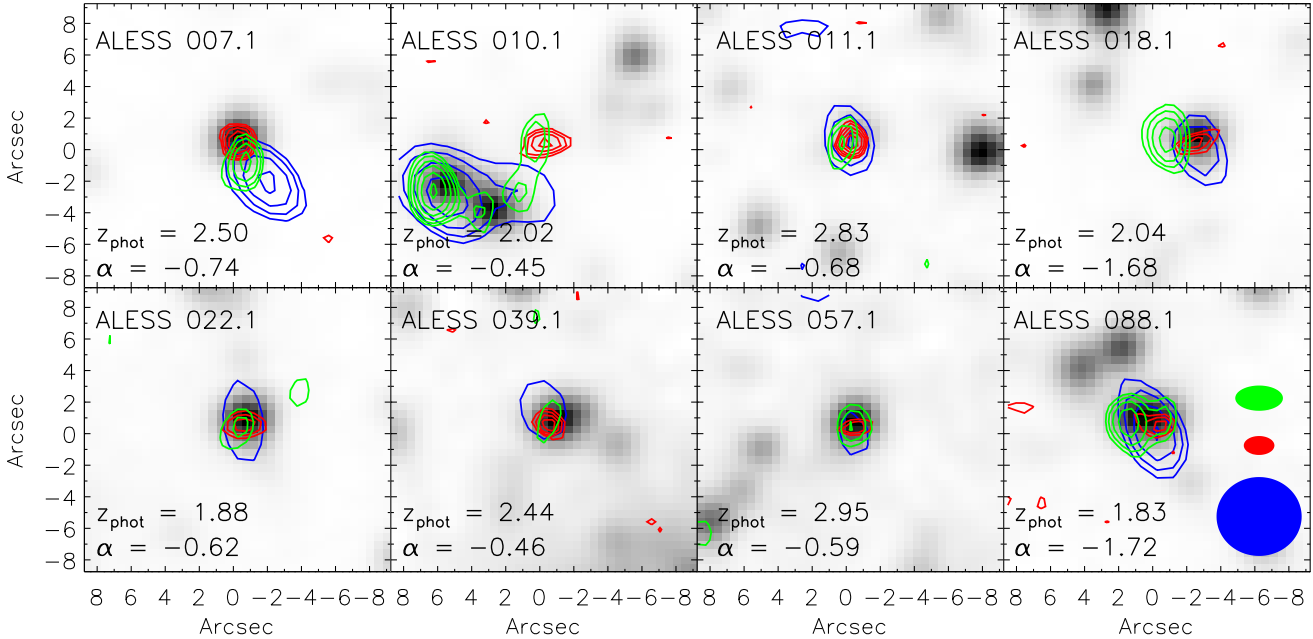}
\vspace{-5mm}
\caption{Postage stamps showing $\sim 19''\times19''$ around eight representative ALESS SMGs with strong radio detections. The underlying greyscale is from \spitzer\ IRAC Channel 1 ($3.6\,\mu$m). The GMRT 610-MHz data are represented with blue contours, the VLA 1.4-GHz data by green contours, and the ALMA 870-$\mu$m data by red contours.  We see good astrometric agreement between all four sets of images. In all cases, contours are spaced at $-3,3,3\sqrt{2}\sigma$ and in steps of $\sqrt{2}\sigma$ thereafter. The field of view of these postage stamps roughly corresponds to the area covered by the LABOCA beam, illustrating the improvement in angular resolution offered by ALMA, and the unambiguity with which the SMG(s) can be precisely located within the LABOCA positional uncertainty. Importantly, we see that much of the strong radio emission seen near ALESS\,010.1 originates from a pair of interlopers at $z_{\rm phot}=0.89\pm0.06$ and $z_{\rm phot}=0.91\pm0.03$, and not from the ALMA SMG at $z_{\rm phot}=2.36\pm0.08$, highlighting the importance of high-resolution ALMA observations to correctly identify the counterpart to the single-dish detection. Typical synthesized beams for the GMRT, VLA and ALMA observations are shown in the bottom-right panel in blue, green and red, respectively.}
\label{fig:postage-stamps}
\end{figure*}

We present ALMA, GMRT and IRAC postage stamps of a representative selection of these SMGs in Figure \ref{fig:postage-stamps}, from which we see the good astrometric agreement between the three sets of data, as well as evidence of multiplicity in the IRAC images.

\begin{table}\label{tab:sample-selection}
\centering
{\small
\caption{ALESS Sample Selection}}
 \begin{tabular}{lcc}
    \hline
    \multicolumn{1}{l}{Selection}&
    \multicolumn{1}{c}{\# sources}&\\
    \multicolumn{1}{l}{}&
    \multicolumn{1}{c}{}&\\
    \hline
    Full ALESS Sample & 131\\
    ALESS {\sc main} Sample & 99\\
    Photometric {\sc main} sample $^a$ & 78\\
    Photometric {\sc main} sample within VLA footprint & 76\\
    VLA-detected sub-sample & 52\\
    GMRT-detected sub-sample$^b$ & 35\\
    \hline
  \end{tabular}
{\small

Notes: $^{a}$From \citet{simpson13}; $^b$all GMRT 610\,MHz-detected SMGs are also detected in the VLA 1.4\,GHz map.}
\end{table}

\section{Analysis and results}

On visual inspection of the radio data, we find that the majority of our sample are unresolved at the resolutions both of the VLA 1.4-GHz and GMRT 610-MHz maps. Since both radio maps are in units of Jy\,beam$^{-1}$, we measure the radio flux densities of these sources by taking the peak pixel value in each map within a 2-arcsec aperture centred on each ALMA ID, and applying a BWS correction from the appropriate pixel in the smearing map (typically $\sim 10$--15\%). For those sources that appear to be extended on scales greater than the VLA synthesized beam, we import postage stamps into \aips\ and measure their flux densities within a polygonal region using the task {\sc blsum}. In total, we detect 52 SMGs at $>3\sigma$ in the VLA map, of which 35 are also detected at $>3\sigma$\ in the GMRT map. All GMRT-detected SMGs have a secure VLA counterpart. For those sources not detected in one or both radio maps, we assign a $3\sigma$\ upper-limit to the flux density using an estimate of the local noise.

In order to investigate the radio-faint SMG population in the ECDFS, we perform a stacking analysis of the non-detections -- stacking both maps at the positions of those ALMA SMGs not detected in either band (Stack\,1), we measure bootstrap median flux densities of $S_{{\rm 1.4 GHz}}=10\pm 1$\,$\mu$Jy and $S_{{\rm 610 MHz}}=38\pm 5$\,$\mu$Jy, while the stack of those sources detected at 1.4\,GHz but not at 610\,MHz (Stack\,2) has median flux densities $S_{{\rm 1.4 GHz}}=30\pm 3$\,$\mu$Jy and $S_{{\rm 610 MHz}}=38\pm 6$\,$\mu$Jy.

\subsection{Radio spectral index measurement}
We use our updated VLA catalogue in conjunction with new GMRT data to measure the radio spectral indices of ALESS SMGs between 610-MHz and 1.4-GHz, which we assume to follow a simple power law, $S_{\nu} \propto \nu^{\alpha}$, where $\alpha$ in any individual galaxy is typically determined by the relative importance of steep-spectrum, non-thermal synchrotron emission ($\alpha \sim -0.8$) versus the flatter-spectrum ($\alpha\sim -0.1$) thermal free-free component which traces ionised H{\sc ii} regions around young, massive stars \citep[][]{condon92}.

\begin{figure*}
\centering
\includegraphics[width=\textwidth]{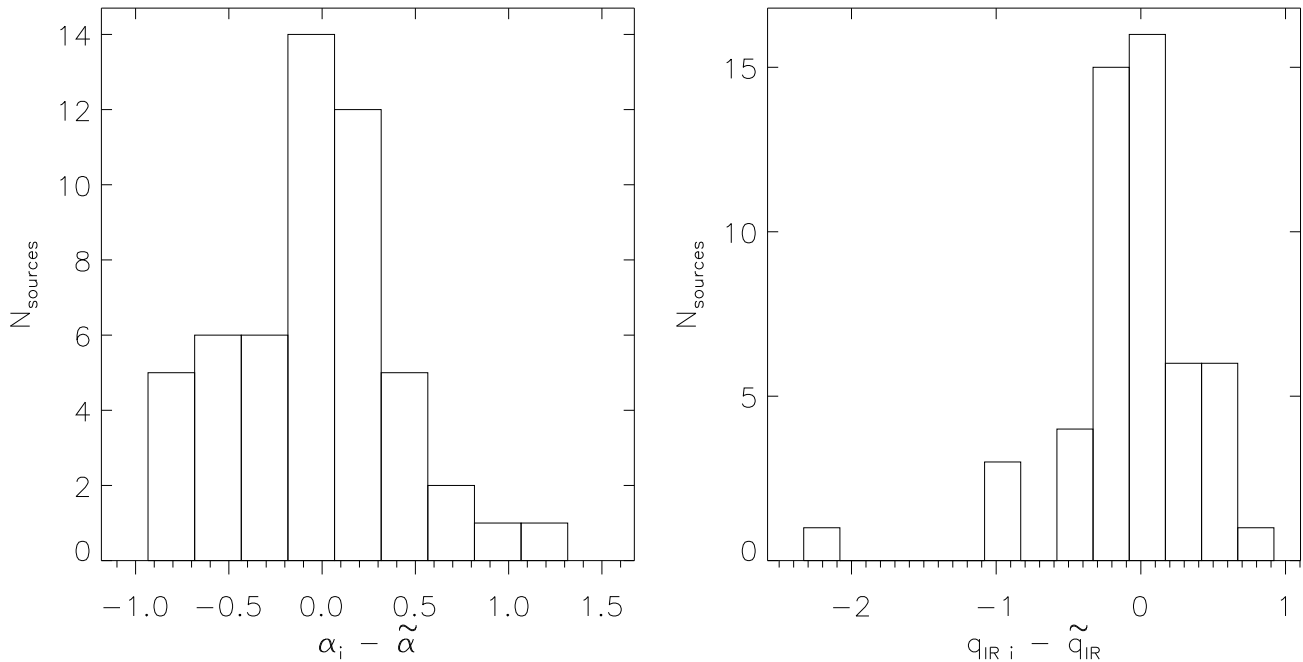}
\vspace{5mm}
\caption{\smallskip Histograms of the absolute deviation from the median in and $\alpha$ and $q_{\rm IR}$, respectively, for the 52 VLA-detected SMGs. The mean absolute deviations in $\alpha$ and $q_{\rm IR}$ are -0.049 and -0.055, respectively. The wide range of radio spectral properties -- which is unaccounted for, in studies which assume $\alpha\sim -0.8$ -- is, due to $K$-correction, partly responsible for the wide spread of $q_{\rm IR}$ measured in this sample.}
\label{fig:mad}
\end{figure*}

In the 52 SMGs for which we have reliable $z_{\rm phot}$ and robust VLA 1.4-GHz detections (including those with limits on their GMRT flux densities), we measure a median spectral index of $\alpha_{\rm all}=-0.79\pm 0.06$ (where the uncertainty given is the standard error, $\sigma/\sqrt{N}$). In Figure\,\ref{fig:mad} we show the distribution of the data ($\alpha_{i}$) around the median ($\tilde{\alpha}$). The Mean Absolute Deviation in these spectral index measurements is ${\rm MAD}=-0.049$. We find that 90\% of the data lie in the range $\alpha=[-1.54,-0.16]$, and that the inter-quartile range is $\alpha=[-1.16,-0.56]$. These results are consistent with the spectral index $\alpha=-0.75\pm0.06$ measured by \citet{ibar10} for a sample of $z\sim 2$ submillimetre sources in the Lockman Hole, and with radio spectra that are, on average, dominated by optically-thin sychrotron emission from star formation. 

We assign spectral indices to those SMGs which are undetected in either map using the flux densities measured from our stacking analysis. For the 24 SMGs with no radio counterpart in either band (Stack\,1), we measure $\alpha_{\rm stack} = -1.57 \pm 0.08$ and $L_{{\rm 1.4 GHz}}=8.3\times 10^{22}$\,W\,Hz$^{-1}$, at the median redshift $\langle z_{\rm phot} \rangle=2.57 \pm 0.06$. For the 17 SMGs with a VLA counterpart but no GMRT counterpart (Stack\,2), we measure $\alpha_{\rm stack} = -0.29 \pm 0.08$, considerably flatter than the median spectral index for the full sample. Using this spectral index, we $K$-correct the stacked 1.4\,GHz flux density to the median redshift of these SMGs, $\langle z_{\rm phot} \rangle=2.30 \pm 0.24$, and measure $L_{{\rm 1.4 GHz}}=1.1\times 10^{24}$\,W\,Hz$^{-1}$. 

\subsection{Far-infrared/radio correlation}

In order to measure the rest-frame 1.4\,GHz radio luminosities ($L_{\rm 1.4 GHz}$) of the ALESS SMGs, we $K$-correct their observed VLA flux densities using measured spectral indices where available (i.e. for those SMGs with secure VLA+GMRT detections), and use limits on $\alpha$ (derived from $3\,\sigma$ GMRT flux limits) to assign upper limits to radio luminosities elsewhere. 

With these measurements of $L_{\rm 1.4\,GHz}$, and using $L_{\rm IR}$ from \citet{swinbank13}, we measure a median $q_{\rm IR}=2.56\pm0.05$ for our full sample of SMGs (${\rm MAD}=-0.055$, and find that 90\% of the data lie between $q_{\rm IR}=[1.67,3.21]$, with inter-quartile range $q_{\rm IR}=[2.42,2.78]$)). Restricting our analysis to the 52 SMGs detected at the VLA gives $q_{\rm IR}=2.58\pm0.07$. Both of these values agree with the measurements of \citet[][who measured $q_{\rm IR}=2.64\pm 0.02$, for a heterogeneous sample of $z<0.1$ star-forming galaxies]{bell03}, \citet[][who measured $q_{\rm IR}=2.67\pm 0.12$ in 591 $z\sim 2$ ULIRGs in the COSMOS field]{sargent10}, and \citet[][who measured $q_{\rm IR}=2.51\pm 0.01$ in a sample of 16 SMGs with high resolution $860\,\mu{\rm m}$ Submillimeter Array imaging, fit with an Arp\,220 template]{barger12}. However, they are higher than the median $q_{\rm IR}=2.40\pm0.12$ reported by \citet{ivison10b} for a sample of IR-bright ($S_{250\,\mu{\rm m}}\gtrsim 20$\,mJy) galaxies out to $z<3.5$, and higher too than $q_{\rm IR}=2.17 \pm 0.19$, as measured by \citet{magnelli10} for a sample of bright ($S_{850 \mu {\rm m}}>5$\,mJy) single-dish submillimetre sources observed in GOODS-N.

Direct comparison of results from these different studies is difficult due to the inhomogeneous way in which the samples have been selected. In both \citet{ivison10b} and \citet{magnelli10}, single-dish submillimetre sources are associated with individual galaxies via the use of statistical techniques, and a combination of mid-IR and radio prior catalogues. Such statistical associations frequently fail to identify a large fraction of submillimetre sources, and those sources that \textit{are} identified may not be representative of the submillimetre-selected population as a whole. For instance, using deep radio and mid-IR imaging, \citet{biggs11} statistically identify probable counterparts to 63\% of the LABOCA-detected SMGs in ECDFS: 45\% with counterparts in both radio and MIPS and 18\% in just a single band ($\sim $\,10\% of these SMGs have {\it several} robust counterparts), leaving at least 37\% still unidentified. In fact, the situation was shown to be even worse when ALMA observations of this sample became available -- in \citet{hodge13} it was shown that the probabilistic radio/mid-IR counterpart identification pinpoints the {\it correct} counterpart to the SMG only $\sim 55$\% of the time, leaving nearly half of the submillimetre sources without identifications, and moreover that $\sim 20$\% of those counterparts claimed as ``robust'' are in fact misidentifications. By directly locating the source of the submillimetre emission with ALMA (without the need for prior radio identification), we create a complete SMG census that includes radio-faint SMGs, and so it is \textit{expected} that the typical $q_{\rm IR}$ should increase relative to previous SMG studies.

\begin{figure*}
\centering
\includegraphics[width=\textwidth]{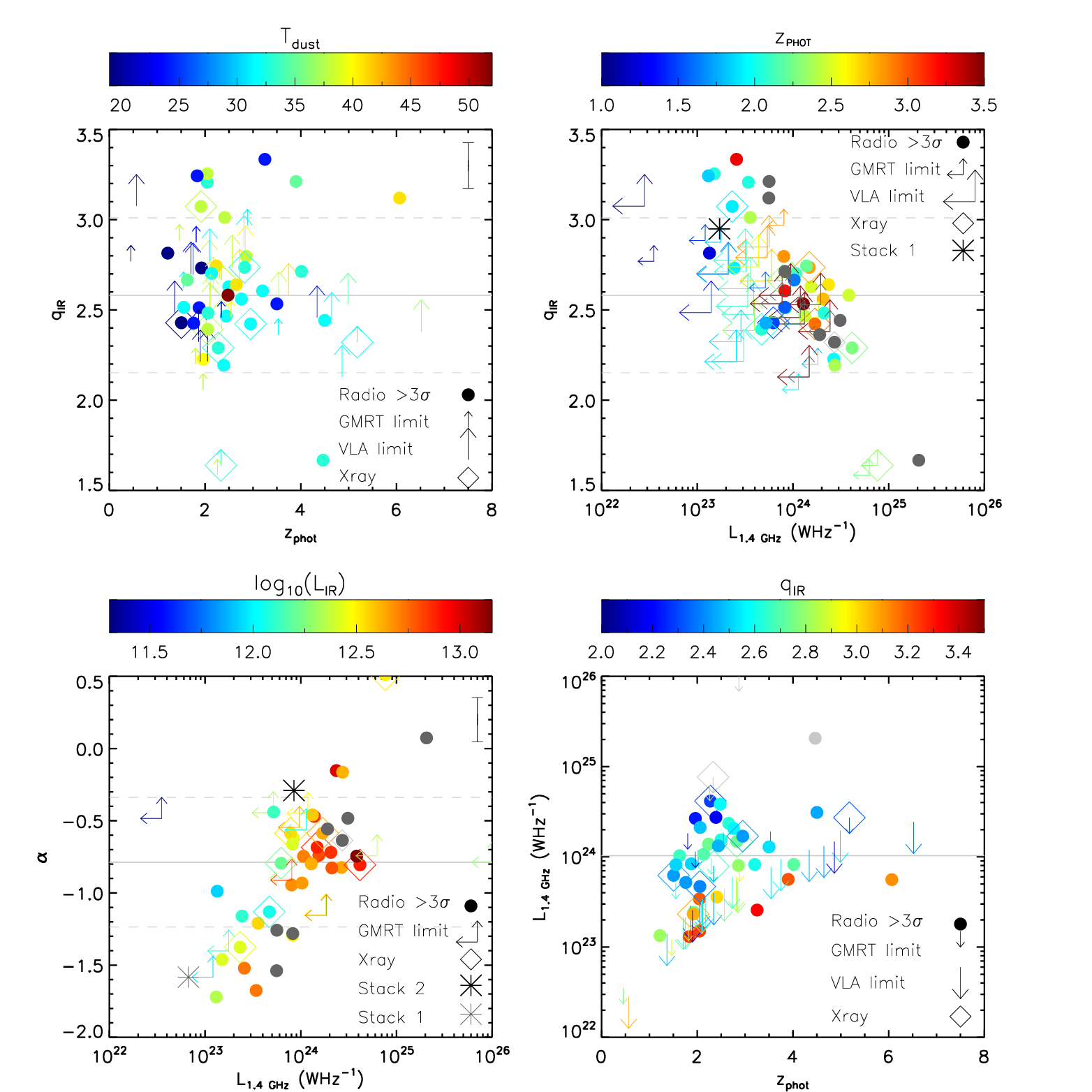}
\vspace{1mm}
\caption{\textit{Top-left:} $q_{\rm IR}$ versus $z_{\rm phot}$, colour-coded by dust temperature $T_{\rm dust}$ for the 76 ALMA-identified SMGs from \citet{hodge13} which lie within the VLA coverage of ECDFS, and for which $z_{\rm phot}$ and $L_{\rm IR}$ are available from \citet{simpson13} and \citet{swinbank13}, respectively. We search for evolution of the form $q_{\rm IR} \propto (1+z)^{\gamma}$, but find that the data are consistent with no evolution: $\gamma=-0.15\pm0.17$. The black bar located in the upper-right corner of the sub-figure shows the typical measurement uncertainty on $q_{\rm IR}$. We see no evidence of a strong correlation between $q_{\rm IR}$ and $T_{\rm dust}$. \textit{Top-right:} $q_{\rm IR}$ versus $L_{\rm 1.4\,GHz}$. For SMGs without a robust radio counterpart, we assign $3\sigma$ limits to $L_{\rm 1.4\,GHz}$ and $q_{\rm IR}$ based on an estimate of the local noise around each source. SMGs with a VLA detection but no GMRT counterpart have unknown spectral indices -- taking the $3\,\sigma$\ GMRT flux limits, we obtain a limit on $\alpha$, which in turn gives upper limit to the $K$-corrected $L_{\rm 1.4\,GHz}$. Included in the plot is a stack of all VLA non-detected sources ('Stack\,1'), where the radio luminosity is calculated by placing the stack at the median redshift, and $K$-correcting using the median spectral index of the sample. \textit{Bottom-left:} Radio spectral index, $\alpha$, versus $L_{\rm 1.4 GHz}$ for the 52 VLA-detected SMGs. Those SMGs with a VLA detection but no GMRT 610\,MHz detection to constrain the spectral index are plotted as $3\sigma$\ limits. We include the result of stacking both radio maps at the positions of ALMA sources with a VLA counterpart but no GMRT counterpart (Stack\,2) and at the positions of ALMA sources with no counterpart in either radio map (Stack\,1, as described above), assuming the stacks lie at the median photometric redshift of each sub-sample and $K$-correcting using the measured spectral index of the stack. The stacks lie below the median $L_{\rm 1.4\,GHz}$, but within the scatter of the data, suggesting that these SMGs are representative of the faint end of the luminosity function. \textit{Bottom-right:} $L_{\rm 1.4\,GHz}$ versus redshift for all 76 SMGs in the ALESS photometric sample. In all four sub-figures, SMGs with properties outside the range of the displayed colour bar are plotted with grey symbols, and SMGs with strong X-ray counterparts from \citet{wang13} are plotted with diamonds.}
\label{fig:fig2}
\end{figure*}

\subsection{ALMA-blank maps}

A surprising finding of our programme of ALMA follow-up observations to the original LESS survey was that 17/88 ``good'' ALMA maps ($\sim 20$\%) contained no ALMA sources above a $4\sigma$ threshold of 2.8\,mJy\,beam$^{-1}$, despite the corresponding LABOCA maps having an average flux $S_{\rm LABOCA}=5.0\pm 0.2$\,mJy \citep{hodge13}. Among the plausible explanations for this phenomenon are: (i) that a significant fraction of the original LABOCA detections are spurious; (ii) that the LABOCA detections are not spurious, but arise from a population of sources that are spatially extended and/or fragment in to multiple components, in which we expect to recover very weak emission within the much smaller ALMA beam; (iii) that incompleteness in source extraction process has resulted in a large number of \textit{bona-fide} SMGs being overlooked in our ALMA maps. Here, we attempt to address this issue by investigating the radio properties of these 17 ``good'', but ALMA-blank fields, using both the VLA and GMRT maps, and the latest photometric redshift catalogue of ECDFS from \citet{simpson13}. 

We fail to find emission in either radio band at the $>3\sigma$ level in 8/17 ALMA-blank pointings (LESS\, 33, 47, 53, 77, 86, 90, 100, 123). Of the remaining 9 fields, 5 (LESS\, 21, 27, 95, 96, 104) have only tentative \spitzer\ MIPS or VLA counterparts to the LABOCA source \citep[with a probability of being misassociations $p>0.05$;][]{biggs11}, while four (LESS\,78, 108, 111, 120) were deemed to have ``robust'' MIPS or radio SMG counterparts in \citet{biggs11}. Of these four ``securely'' identified submillimetre sources, one (LESS\,78) lies outside the footprint of the MUSYC near-IR survey, and hence is not in the photometric redshift catalogue of \citet{simpson13}. Of the remaining three ALMA-blank LABOCA sources, one (LESS\,108) shows strong ($>10\sigma$) VLA and GMRT emission extending over $\sim48''^2$ ($\sim 24$ ALMA beam areas), which appears to be associated with a foreground ($z_{\rm phot}=0.19\pm0.06$) disk galaxy. If the LABOCA flux of $9.3\pm1.2$\,mJy reported in \citet{weiss09} is extended over a similarly large area, then we expect a peak ALMA flux density of $\sim 0.4$\,mJy\,beam$^{-1}$, approximately $1\times\sigma$ in the good maps. Thus the ALMA non-detection of LESS\,108 is unsurprising. The remaining two ALMA-blank LABOCA sources (LESS\,111, 120) each have a single, plausible $\sim4\sigma$ radio counterpart within $\sim8''$ of the LABOCA centroid, lying at $z_{\rm phot}=2.59\pm0.06$ and $z_{\rm phot}=1.51\pm0.17$, respectively. However in addition, these two fields also feature 4 and 7 MUSYC or IRAC galaxies, respectively; given that ALMA sees no evidence of strong, compact 870-$\mu$m emission in either field, we conclude that the measured LABOCA fluxes within these fields ($S_{\rm LABOCA}=5.2\pm1.3$\,mJy and $S_{\rm LABOCA}=4.9\pm1.3$\,mJy) are most likely the sum of the fluxes of these individual galaxies, none of which individually may lie above the ALMA detection threshold. LESS\,108 is clearly misidentified as a radio-detected SMG; if the \citet{biggs11} radio counterparts to LESS\,78, 111 \& 120 are similarly misidentified, then the fraction of ``robust'' counterparts in the ALMA-blank maps which are misidentifications is 4/17 ($\sim24$\%), comparable to the fraction of radio-misidentified counterparts reported in \citet{hodge13} for the ALESS {\sc main} sample.

 We note that 13/17 ALMA-blank LABOCA sources also lack strong radio counterparts, and thus, we conclude that there is no reason to believe our ALMA source extraction process has missed a significant number of \textit{bona-fide} luminous $z\sim 2$ SMGs. We believe that intrinsic properties of these LABOCA sources (i.e. extreme multiplicity and/or very extended emission) are the reasons why they are not detected in our ALMA maps.  

\section{Discussion and conclusions}
\subsection{Radio spectral properties}

While the majority of SMGs in our sample have measured spectral indices within the range defined by free-free ($\alpha \sim -0.1$) and synchrotron emission ($\alpha \sim -0.8$), a  minority have spectral indices outside this range. Of the (biased) subset of 35 SMGs with robust counterparts in both radio maps and $z_{\rm phot}$, 12 are ultra-steep spectrum (USS; $\alpha < -1.0$) sources lying at a median $z=2.4\pm0.4$, while two SMGs (at $z=2.33$ and $z=4.47$) have inverted ($\alpha > 0$) radio spectra. Such diversity in the radio spectral properties of galaxies has been seen before, for example by \citet{murphy13}, who in a study of the radio properties of 31 major mergers demonstrated a correlation between $\alpha$\ and the merger phase, wherein early-stage mergers (with distinct galaxy nuclei but interpenetrating, strongly synchrotron-emitting magnetic fields forming a ``bridge'' between nuclei) have the steepest spectra ($\alpha<-1.0$), and late-stage (i.e. compact, fully coalesced) major mergers have flatter spectra ($\alpha \gtrsim 0.0$) due to higher free-free optical depths. Ultra-steep spectrum radio emission is also seen around high-redshift radio galaxies (HzRGs; indeed this is how they are detected), which may be due to a combination of observational effects \citep[e.g. redshifting of a concave radio spectrum which flattens at higher frequencies:][]{afonso11}, or may be intrinsic, i.e. linked to the expansion of collimated jets in dense environments \citep{bryant09}, or due to compact nuclear activity \citep{argo13}. 

In order to determine whether our USS SMGs are associated with early-stage merger activity, we search for evidence that they lie in over-dense regions of ECDFS; if merger activity is responsible for these steep radio spectra, then we may expect to find that these SMGs have, on average, more close companions than do those SMGs whose radio spectra are more typical of fully coalesced starbursts. We compute the redshift distribution, $N(z)$, for all IRAC-detected galaxies within the ALMA primary beam centred on each of the USS SMGs, which we compare with the aggregate $N(z)$ for all IRAC-detected galaxies within the primary beam of all ALMA SMGs. From this analysis, we find weak evidence ($\sim2\sigma$) of an excess of 0.5 IRAC sources within the ALMA primary beam between $2.0<z<3.5$ with respect to the density of IRAC sources around non-USS SMGs.

\subsection{Far-infrared/radio correlation}

Our median $q_{\rm IR}=2.56\pm 0.05$ is higher than previous measurements obtained for SMG samples \citep[e.g.][]{ivison10b, magnelli10}, although we note that previous SMG samples have typically been biased towards sources with strong radio emission via the reliance on radio imaging to obtain counterpart IDs. Our latest measurement of $q_{\rm IR}$ is in good agreement with the value measured in local star-forming galaxies \citep{bell03}.

In Figure \ref{fig:fig2} we plot $q_{\rm IR}$ versus redshift for our sample of 76 ALESS SMGs. We search for evolution in $q_{\rm IR}$, but find no compelling evidence for any such evolution: we fit $q_{\rm IR} \propto (1+z)^{\gamma}$, where $\gamma=-0.15\pm0.17$ \citep[cf. the weak evolution $\gamma = -0.26 \pm 0.07$ measured by][]{ivison10b}.  If we further restrict our analysis to the 35 SMGs with secure ($>3\sigma$) detections in both radio maps and $z_{\rm phot}$ (i.e. those SMGs for which we can directly measure $\alpha$), then we measure $\gamma=-0.01\pm0.09$. However, we caution that this sample is more biased than the previous fit.  

The lack of any measurable evolution in $q_{\rm IR}$ with redshift (both within our sample, and between our sample and local results) is at slight odds with theoretical predictions, which suggest that $q_{\rm IR}$ should begin to increase at greater look-back times (i.e, $z \gtrsim 3$), where the warmer CMB temperature should lead to increased Inverse Compton-cooling of cosmic-ray electrons with redshift, which in turn should suppress the non-thermal synchrotron emission from normal star-forming galaxies \citep[e.g.][]{murphy09}.  That we measure no such evolution most likely suggests that the magnetic fields in these SMGs are large enough (i.e., $\gtrsim 50\,\mu$G) to combat this effect.  

However, it is also possible that other physical processes may be at work, which conspire to offset the increased effect of IC-cooling at high redshift and preserve the local FIRRC.  A number of candidate processes are discussed in \citet{murphy09} as the result of the high ISM density of SMGs, including: (i) an increase in the  production of secondary electrons arising from an increase in proton-gas collisions \citep[also see][]{lacki10};  (ii) an increase in the acceleration efficiencies of cosmic-ray electrons due to shortening the adiabatic expansion of supernova remnants.  In the case where the SMGs are in fact major mergers, a significant amount of excess radio emission may arise from tidal bridges and tails that are not associated with star formation \citep{murphy13}.  Conspiracy to maintain the FIRRC at high-redshift could also be achieved due to variations in the dust properties of galaxies in the early universe, including the chemical composition of dust, the dust-to-gas mass ratio, and the size distribution of grains.  While possible, \citet{draine09} argue that these properties do not appear to vary significantly, at least between local star-forming galaxies and for high-redshift QSOs. In Figure\,3, we see at a given redshift no evidence of a correlation between $q_{\rm IR}$ and the dust temperature, $T_{\rm dust}$, measured in \citep{swinbank13}; however the measured dust temperatures do systematically increase with redshift. If this increase in $T_{\rm dust}$ with $z$ correlates with a decrease in SMG size with increasing redshift, then the associated increase in energy density with redshift would provide an additional resilience against CMB losses.

\subsection{Identification of candidate AGN}\label{sect:agn}

X-ray studies \citep[e.g.][]{alexander05, pope06} have been successful in identifying signs of AGN activity in some SMGs. However it is known that even the deepest such studies have sensitivity limits that preclude the detection of the most obscured, Compton-thick AGN, whose intrinsic X-ray spectra are heavily attenuated by the intervening column of dust and gas ($N_{\rm H} \gtrsim 10^{24}$\,cm$^{-2}$). Mid-IR studies \citep[e.g.][]{valiante07, pope08} have provided one means of (indirectly) detecting heavily obscured AGN, by tracing the dust heated to $T_{\rm dust} \sim 200$--$1000$\,K via X-ray absorption, however intense star formation activity can dominate the entire IR SEDs of galaxies, which makes the task of decomposing SEDs into separate contributions arising due to star formation and AGN non-trivial \citep{mullaney11}.

Radio emission provides an extinction-free means of identifying AGN, and has traditionally divided AGN into two groups: (i) radio-loud AGN ($L_{\rm 1.4 GHz} \gtrsim 10^{24}$\,W\,Hz$^{-1}$), which exhibit radio jets and lobes on scales of several kpc \citep{yun01}; and (ii) radio-quiet AGN, which produce weaker, flat-spectrum radio emission that is confined within a compact, pc-scale core \citep{blundell07}. \citet{delmoro13} use a joint analysis in the IR and radio bands to isolate such ``radio excess'' sources, which deviate significantly from the FIRRC relation, and perform multi-band SED fits to constrain the influence of any mid-IR AGN component. Here, we briefly perform a similar analysis to identify candidate radio-excess AGN hosts -- we find four SMGs whose $q_{\rm IR}$ values are $> 2\sigma$ lower than the sample median ($q_{\rm IR} < 1.7$). Of these four SMGs, one (ALESS 066.1) has already been classified as an AGN host based on its X-ray spectral properties \citep{wang13}. Its inverted radio spectrum ($\alpha > 0.51$, based on a $3\,\sigma$\ GMRT flux limit) and high radio luminosity ($L_{\rm 1.4 GHz}=7.6\times 10^{24}$\,W\,Hz$^{-1}$) are consistent with compact nuclear activity. Of the remaining three $q_{\rm IR} < 1.7$ SMGs, one (ALESS 014.1) has a flat radio spectrum ($\alpha>0.07$) and $L_{\rm 1.4 GHz} = 2.0 \times 10^{25}$\,W\,Hz$^{-1}$, while the other two (ALESS 094.1 \& ALESS 118.1, both of which have VLA but not GMRT detections) have spectral index \textit{limits} that are consistent with star formation ($\alpha > -0.78$) and radio luminosities $L_{\rm 1.4 GHz} < 1.0\times 10^{26}$\,W\,Hz$^{-1}$ and $L_{\rm 1.4 GHz} < 6.3\times 10^{24}$\,W\,Hz$^{-1}$. Taken at face value such high radio luminosities are strongly suggestive of powerful AGN activity, however we caution that three of these candidate AGN-hosts are not detected in the 610\,MHz GMRT map. By assigning $3\sigma$\ upper limits to their GMRT flux densities, we impose a steep spectrum and strong $K$-correction which will naturally boost $L_{\rm 1.4 GHz}$ and lower $q_{\rm IR}$. We therefore caution that although $q_{\rm IR}$ is a useful metric for the identification of candidate AGN, the implications of using this metric are unclear if the spectral index, $\alpha$, is not well constrained (as it is in ALESS 014.1).

\subsection{Trends in $\alpha$\ and $q_{\rm IR}$ in high-redshift ULIRGs}

\begin{figure*}
\centering
\includegraphics[width=\textwidth]{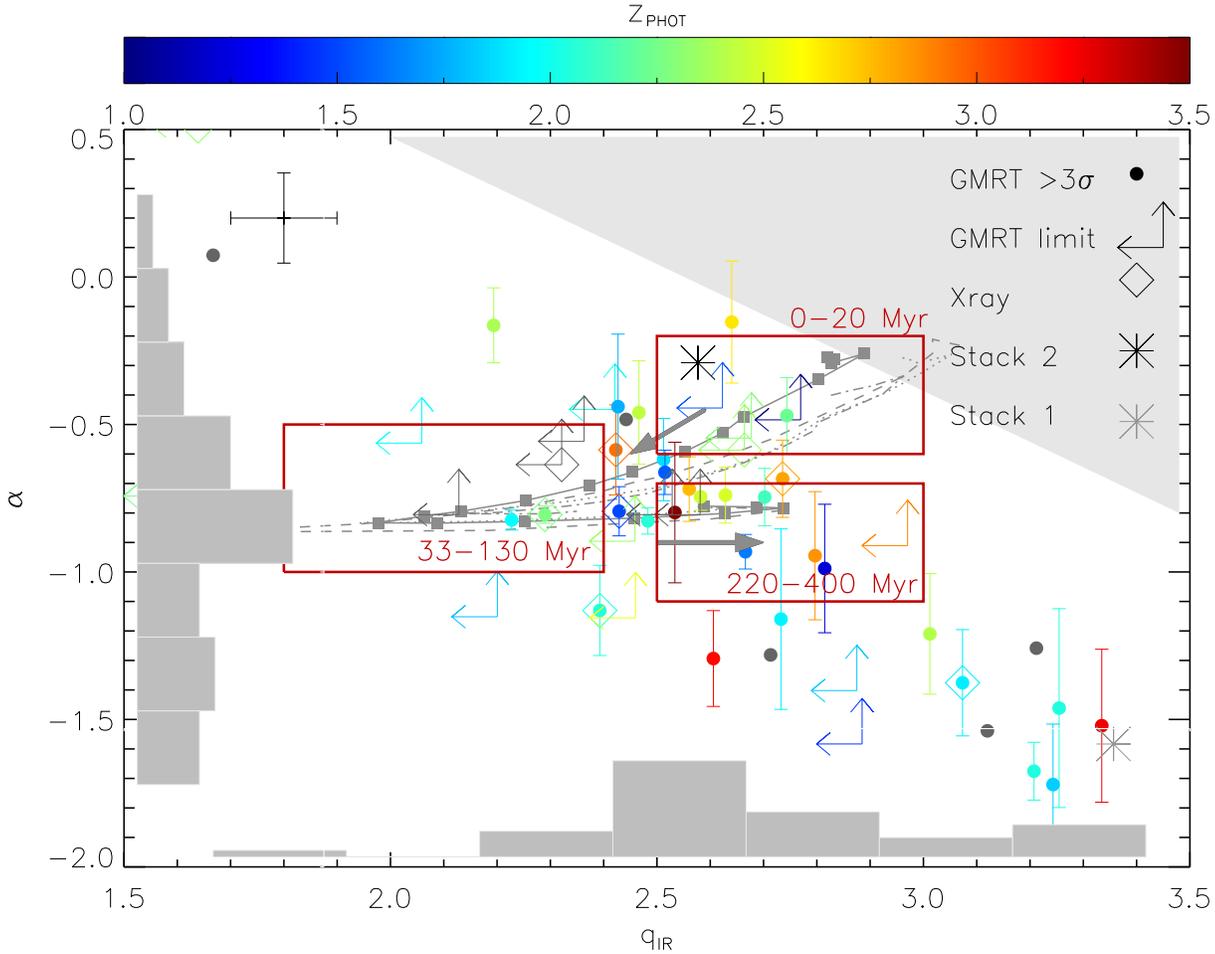}
\vspace{3mm}
\caption{Radio 610\,MHz/1.4\,GHz spectral index, $\alpha$, versus far-infrared/radio correlation parameter, $q_{\rm IR}$ for the 52 VLA-detected SMGs in ECDFS (histograms of both quantities illustrate the widths of the distributions). Galaxies are colour-coded by redshift between $1<z<3.5$, with all galaxies outside this range plotted in light grey. We plot $3\sigma$\ limits for those SMGs with a VLA detection but no corresponding GMRT detection to constrain $\alpha$. Galaxies with a confirmed X-ray detection are over-plotted with colour-coded diamonds. In the top-left we show typical error bars on both quantities. The large black and grey stars show the locations of stacks of (i) SMGs with a VLA detection but no GMRT counterpart (Stack\,2), and (ii) SMGs with no detection in either radio map (Stack\,1), respectively. The shaded region at the top-right highlights a selection boundary (see text for details). Lastly, we plot example tracks from the models of \citet{bressan02}, along which galaxies are expected to evolve in the directions indicated, and isolate regions where we expect to find old, middle-aged, and young galaxies. The model tracks run from $\log(t_{\rm yr})=6.3$--$8.6$ in 21 logarithmically-spaced steps (marked with filled grey squares). From the far-IR properties reported in \citet{swinbank13}, we measure weak correlations between both $M_{\star}$ and $M_{\rm dust}$ and predicted age as expected for this evolution.}
\label{fig:alpha-v-qir}
\end{figure*}

In Figure \ref{fig:alpha-v-qir}, we plot $\alpha$\ and $q_{\rm IR}$ for the ALESS SMGs, and over-plot model tracks from \citet{bressan02} for evolving starburst galaxies. Histograms on the $x$ and $y$ axes highlight the breadth of the distributions in $\alpha$\ and $q_{\rm IR}$. We see in the data an apparent linear correlation between $\alpha$ and $q_{\rm IR}$, and investigate whether this correlation is real or the result of selection effects by generating a catalogue of 500,000 mock galaxies with $L_{\rm IR}$, $\alpha$\ and $q_{\rm IR}$  drawn randomly from the data, with random redshifts in the range $1.5<z<3.5$, and computing the corresponding flux densities at 870-$\mu$m, 1.4-GHz and 610-MHz. We classify our mock galaxies as ``observable'' if both the 870\,$\mu$m and 1.4\,GHz flux densities are above the survey sensitivity limits, and ``unobservable'' if one or both flux densities lie below the corresponding threshold. Next, we bin the ``observable'' galaxies in $q_{\rm IR}$ with bin widths $\Delta{\rm q}_{\rm IR}=0.1$ and determine the $\pm5\sigma$\ range of $\alpha$\ in each bin. Hence we determine a $5\sigma$\ selection boundary (shaded in light grey) precluding the detection of flat/inverted spectrum sources that also have a high far-IR/radio flux ratio. We see no evidence of a selection boundary within the plot range affecting our ability to detect steep spectrum SMGs with low $q_{\rm IR}$. From this, we conclude that the lack of ultra steep-spectrum sources with low $q_{\rm IR}$ in our sample is real. 

In models of isolated, obscured starbursting galaxies \citep[e.g.][]{bressan02}, the relative strengths of radio and far-infrared emission are predicted to be a sensitive function of age, such that over its lifetime, a starbursting galaxy will trace out a unique track in a plot of $\alpha$\ versus $q_{\rm IR}$. At early times, heating of dust by massive stars is thought to elevate $L_{\rm IR}$ before the first supernovae occur. The radio spectrum of such a young starburst is hence likely to be dominated by the flatter-component free-free emission, and $q_{\rm IR}$ is elevated. As the starburst ages, non-thermal radio emission from supernova-accelerated cosmic ray electrons rapidly begins to dominate, steepening the radio spectrum, boosting $L_{\rm 1.4 GHz}$, and lowering $q_{\rm IR}$. As the starburst activity finally draws to a halt, both $L_{\rm IR}$ and $L_{\rm 1.4 GHz}$ decline over time such that $q_{\rm IR}$ and $\alpha$ approach asymptotic values. 

\citet{bressan02} provide model tracks of $\alpha$\ and $q_{\rm IR}$ for evolving starburst galaxies, which we test against our data by segregating the galaxies in Figure \ref{fig:alpha-v-qir} into three ``age'' groups, representing the ``old'' (220--400\,Myr), ``middle-aged'' (33--130\,Myr) and ``young'' (0--20\,Myr) ends of the model tracks. We look for correlations of age with the star formation rates (SFRs), dust masses ($M_{\rm dust}$), dust temperatures ($T_{\rm dust}$) and stellar masses ($M_{\star}$) measured by \citet{swinbank13}. We find no evidence of significant correlations between SFR or $T_{\rm dust}$ and location on the evolutionary tracks, but do find evidence of weak correlations between $M_{\star}$ and age, and between $M_{\rm dust}$ and age (see Table\,2). These correlations are weak due to the small number of SMGs in each ``age'' bin, but it is suggestive that the correlations appear to run in the sense that they are expected to, i.e. galaxies at the more ``evolved'' ends of the tracks appear to have higher $M_{\star}$ and $M_{\rm dust}$.

At the bottom-right corner of the plot, the USS SMGs depart significantly from the \citet{bressan02} model tracks. If these galaxies are indeed on-going, early-stage major merger events, then they may represent a population of proto-starburst galaxies which have not yet reached the beginning of the \citet{bressan02} starburst evolutionary tracks. However, we find that their high values of $q_{\rm IR}$ are not due to those SMGs being abnormally bright at infrared wavelengths ($\langle L_{\rm IR\,USS}\rangle=(3.6\pm 0.7)\times 10^{12}$\,L$_{\odot}$, compared to $\langle L_{\rm IR}\rangle =(3.2\pm 0.3)\times 10^{12}$\,L$_{\odot}$ for the non steep-spectrum ALESS {\sc main} sample), but rather are the result of them being comparatively \textit{faint} in the radio ($\langle L_{\rm 1.\,GHz,USS}\rangle=(3.6\pm 0.3)\times 10^{23}$\,W\,Hz$^{-1}$, compared to $\langle L_{\rm 1.4\,GHz}\rangle = (8.0\pm 1.4)\times 10^{23}$\,W\,Hz$^{-1}$ in the {\sc main} SMGs). The measurement of a suppressed radio luminosity in the USS SMGs is at odds with the proposed mechanism for steepening their radio spectra in the first place \citep[\S\,4.1;][]{murphy13}. An alternative hypothesis is that some of the USS sources may in fact be low-luminosity examples of the ``compact steep spectrum'' class of objects \citep{odea98}, whose radio emission is thought to arise primarily in frustrated radio jets whose propagation away from the central black hole is truncated by interactions with dense gas in their environments. Differentiating between these two scenarios is beyond the scope of the present work, due to the need for radio maps of kiloparsec-scale resolution to be able to detect asymmetries in the radio surface brightness profiles of these USS sources, which would be the ``smoking gun'' of frustrated jet activity.

\begin{table}
\centering
{\small
\caption{Properties of SMGs selected in different age bins}}
 \begin{tabular}{lcccc}
    \hline
    \multicolumn{1}{l}{Age bin}&
    \multicolumn{1}{c}{\# SMGs}&
    \multicolumn{1}{c}{$M_{\star}^a$}&
    \multicolumn{1}{c}{$M_{\rm dust}^a$}\\
    \multicolumn{1}{l}{(Myr)}&
    \multicolumn{1}{c}{}&
    \multicolumn{1}{c}{($\times 10^{11}$\,M$_{\odot}$)}&
    \multicolumn{1}{c}{($\times 10^8$\,M$_{\odot}$)}&\\
    \hline
    0--20 & 5 & $0.84\pm0.27$ & $2.54\pm0.31$\\
    33--130 & 6 & $1.02\pm0.28$ & $5.18\pm0.69$\\
    220--400 & 9 & $2.12\pm0.64$ & $5.24\pm0.25$\\
    \hline
  \end{tabular}\label{tab:age}
{\small

Notes: $^a$From \citet{swinbank13}. Uncertainties are given by the standard error, $\sigma/\sqrt{N}$, where $N$ is the number of SMGs in each bin.}
\end{table}

\section{Conclusions}

We have presented the results of a study into the radio properties of ALMA-identified SMGs from the LABOCA 870\,$\mu$m survey of the ECDFS \citep{weiss09, hodge13}. Bringing together high-angular resolution ALMA 870\,$\mu$m observations, publicly available VLA 1.4\,GHz imaging \citep{miller13} and new, deep 610\,MHz imaging from GMRT, we have been able for the first time both to (i) obtain unbiased identification of counterparts to the submillimetre source without using prior radio/mid-IR observations, and (ii) $K$-correct the observed radio flux densities and measure radio luminosities ($L_{\rm 1.4GHz}$) using an individual spectral index ($\alpha$) for each SMG. The radio properties of the ALESS SMGs are presented in Table\,3.

For the full sample of 76 SMGs covered by the VLA observations, we measure a median $\alpha = -0.79 \pm 0.06$, consistent with previous studies of SMGs \citep[e.g.][]{ibar10}, but with a number of sources with peculiarly flat/inverted ($\alpha \gtrsim 0.0$) or steep ($\alpha < -1.0$) radio spectra, and measure a median logarithmic far-IR/radio flux ratio $q_{\rm IR}=2.56\pm 0.05$. Finally, we have plotted the measured $\alpha$\ and $q_{\rm IR}$ for each VLA-detected galaxy, finding that the majority lie close to the evolutionary tracks of \citet{bressan02}, and that there is tentative evidence of a correlation between progress along the evolutionary track and both $M_{\star}$ and $M_{\rm dust}$.

Together, these findings highlight the potential for targeted ALMA and radio interferometric follow-up observations of SMGs detected in wide area submillimetre surveys to act as a laboratory for testing models of galaxy evolution at $z\gtrsim 2$.

\section*{Acknowledgments} 
We thank the editor and anonymous referee for their constructive comments which have led to the improvement of this paper. This paper makes use of the following ALMA data: ADS/JAO.ALMA\#2011.0.00294.S. ALMA is a partnership of ESO (representing its member states), NSF (USA) and NINS (Japan), together with NRC (Canada) and NSC and ASIAA (Taiwan), in cooperation with the Republic of Chile. The Joint ALMA Observatory is operated by ESO, AUI/NRAO and NAOJ. We thank the staff of the GMRT who made these observations possible. GMRT is run by the National Centre for Radio Astrophysics of the Tata Institute of Fundamental Research. This research also made use of data taken as part of the HerMES Key Programme from the SPIRE instrument team, ESAC scientists and a mission scientist. \textit{Herschel} is an ESA space observatory with science instruments provided by European-led Principal Investigator consortia and with important participation from NASA. All of the ALMA, Herschel, VLA and \textit{Spitzer} data employed in this analysis are available through the ESO, \textit{Herschel}, VLA and \textit{Spitzer} archives. APT and IRS acknowledge support from STFC (ST/I001573/1). AMS gratefully acknowledges an STFC Advanced Fellowship through grant number ST/H005234/1. IRS also acknowledges a Leverhulme Fellowship, the ERC Advanced Investigator Programme {\sc dustygal} (\#321334) and a Royal Society/Wolfson merit award. RJI acknowledges support from the ERC in the form of Advanced Grant {\sc cosmicism}. JMS and ALRD acknowledge the support of STFC studentships (ST/J501013/1 and ST/F007299/1, respectively). EI acknowledges funding from CONICYT/FONDECYT postdoctoral project N$^\circ$:3130504. The Dark Cosmology Centre is funded by The Danish National Research Foundation. The National Radio Astronomy Observatory is a facility of the National Science Foundation operated under cooperative agreement by Associated Universities, Inc.

\bibliography{references}

\begin{thebibliography}{}

\bibitem[\protect\citeauthoryear{{Afonso} et~al.}{{Afonso}
  et~al.}{2011}]{afonso11}
{Afonso} J. et~al., 2011, \apj, 743, 122

\bibitem[\protect\citeauthoryear{{Alexander} et~al.}{{Alexander}
  et~al.}{2005}]{alexander05}
{Alexander} D.~M., {Bauer} F.~E., {Chapman} S.~C., {Smail} I., {Blain} A.~W.,
  {Brandt} W.~N.,  {Ivison} R.~J., 2005, \apj, 632, 736

\bibitem[\protect\citeauthoryear{{Appleton} et~al.}{{Appleton}
  et~al.}{2004}]{appleton04}
{Appleton} P.~N. et~al., 2004, \apjs, 154, 147

\bibitem[\protect\citeauthoryear{{Argo} et~al.}{{Argo} et~al.}{2013}]{argo13}
{Argo} M.~K., {Paragi} Z., {R{\"o}ttgering} H., {Kl{\"o}ckner} H.-R., {Miley}
  G.,  {Mahmud} M., 2013, \mnras, 431, L58

\bibitem[\protect\citeauthoryear{{Barger} et~al.}{{Barger}
  et~al.}{2012}]{barger12}
{Barger} A.~J., {Wang} W.-H., {Cowie} L.~L., {Owen} F.~N., {Chen} C.-C.,
  {Williams} J.~P., 2012, \apj, 761, 89

\bibitem[\protect\citeauthoryear{{Bell}}{{Bell}}{2003}]{bell03}
{Bell} E.~F., 2003, \apj, 586, 794

\bibitem[\protect\citeauthoryear{{Biggs} et~al.}{{Biggs}
  et~al.}{2011}]{biggs11}
{Biggs} A.~D. et~al., 2011, \mnras, 413, 2314

\bibitem[\protect\citeauthoryear{{Blain} et~al.}{{Blain}
  et~al.}{2002}]{blain02}
{Blain} A.~W., {Smail} I., {Ivison} R.~J., {Kneib} J.-P.,  {Frayer} D.~T.,
  2002, \physrep, 369, 111

\bibitem[\protect\citeauthoryear{{Blundell} \& {Kuncic}}{{Blundell} \&
  {Kuncic}}{2007}]{blundell07}
{Blundell} K.~M.,  {Kuncic} Z., 2007, \apjl, 668, L103

\bibitem[\protect\citeauthoryear{{Bolzonella}, {Miralles} \&
  {Pell{\'o}}}{{Bolzonella} et~al.}{2000}]{bolzonella00}
{Bolzonella} M., {Miralles} J.-M.,  {Pell{\'o}} R., 2000, \aap, 363, 476

\bibitem[\protect\citeauthoryear{{Bourne} et~al.}{{Bourne}
  et~al.}{2011}]{bourne11}
{Bourne} N., {Dunne} L., {Ivison} R.~J., {Maddox} S.~J., {Dickinson} M.,
  {Frayer} D.~T., 2011, \mnras, 410, 1155

\bibitem[\protect\citeauthoryear{{Boyle} et~al.}{{Boyle}
  et~al.}{2007}]{boyle07}
{Boyle} B.~J., {Cornwell} T.~J., {Middelberg} E., {Norris} R.~P., {Appleton}
  P.~N.,  {Smail} I., 2007, \mnras, 376, 1182

\bibitem[\protect\citeauthoryear{{Bressan}, {Silva} \& {Granato}}{{Bressan}
  et~al.}{2002}]{bressan02}
{Bressan} A., {Silva} L.,  {Granato} G.~L., 2002, \aap, 392, 377

\bibitem[\protect\citeauthoryear{{Bridle} \& {Schwab}}{{Bridle} \&
  {Schwab}}{1989}]{bridle89}
{Bridle} A.~H.,  {Schwab} F.~R., 1989, in Astronomical Society of the Pacific
  Conference Series, Vol.~6, {Perley} R.~A., {Schwab} F.~R.,  {Bridle} A.~H.,
  eds, Synthesis Imaging in Radio Astronomy, p. 247

\bibitem[\protect\citeauthoryear{{Bryant} et~al.}{{Bryant}
  et~al.}{2009}]{bryant09}
{Bryant} J.~J., {Johnston} H.~M., {Broderick} J.~W., {Hunstead} R.~W., {De
  Breuck} C.,  {Gaensler} B.~M., 2009, \mnras, 395, 1099

\bibitem[\protect\citeauthoryear{{Carilli} \& {Yun}}{{Carilli} \&
  {Yun}}{1999}]{carilli99}
{Carilli} C.~L.,  {Yun} M.~S., 1999, \apjl, 513, L13

\bibitem[\protect\citeauthoryear{{Casey} et~al.}{{Casey}
  et~al.}{2013}]{casey13}
{Casey} C.~M. et~al., 2013, \mnras

\bibitem[\protect\citeauthoryear{{Condon}}{{Condon}}{1992}]{condon92}
{Condon} J.~J., 1992, \araa, 30, 575

\bibitem[\protect\citeauthoryear{{Damen} et~al.}{{Damen}
  et~al.}{2011}]{damen11}
{Damen} M. et~al., 2011, \apj, 727, 1

\bibitem[\protect\citeauthoryear{{de Jong} et~al.}{{de Jong}
  et~al.}{1985}]{dejong85}
{de Jong} T., {Klein} U., {Wielebinski} R.,  {Wunderlich} E., 1985, \aap, 147,
  L6

\bibitem[\protect\citeauthoryear{{Del Moro} et~al.}{{Del Moro}
  et~al.}{2013}]{delmoro13}
{Del Moro} A. et~al., 2013, \aap, 549, A59

\bibitem[\protect\citeauthoryear{{Donley} et~al.}{{Donley}
  et~al.}{2005}]{donley05}
{Donley} J.~L., {Rieke} G.~H., {Rigby} J.~R.,  {P{\'e}rez-Gonz{\'a}lez} P.~G.,
  2005, \apj, 634, 169

\bibitem[\protect\citeauthoryear{{Draine}}{{Draine}}{2009}]{draine09}
{Draine} B.~T., 2009, in Astronomical Society of the Pacific Conference Series,
  Vol. 414, {Henning} T., {Gr{\"u}n} E.,  {Steinacker} J., eds, Cosmic Dust -
  Near and Far, p. 453

\bibitem[\protect\citeauthoryear{{Garrett}}{{Garrett}}{2002}]{garrett02}
{Garrett} M.~A., 2002, \aap, 384, L19

\bibitem[\protect\citeauthoryear{{Gawiser} et~al.}{{Gawiser}
  et~al.}{2006}]{gawiser06}
{Gawiser} E. et~al., 2006, \apjs, 162, 1

\bibitem[\protect\citeauthoryear{{Griffin} et~al.}{{Griffin}
  et~al.}{2010}]{griffin10}
{Griffin} M.~J. et~al., 2010, \aap, 518, L3

\bibitem[\protect\citeauthoryear{{Hodge} et~al.}{{Hodge}
  et~al.}{2013}]{hodge13}
{Hodge} J.~A. et~al., 2013, \apj, 768, 91

\bibitem[\protect\citeauthoryear{{Hsieh} et~al.}{{Hsieh}
  et~al.}{2012}]{hsieh12}
{Hsieh} B.-C., {Wang} W.-H., {Hsieh} C.-C., {Lin} L., {Yan} H., {Lim} J.,  {Ho}
  P.~T.~P., 2012, \apjs, 203, 23

\bibitem[\protect\citeauthoryear{{Hughes} et~al.}{{Hughes}
  et~al.}{1998}]{hughes98}
{Hughes} D.~H. et~al., 1998, \nat, 394, 241

\bibitem[\protect\citeauthoryear{{Ibar} et~al.}{{Ibar} et~al.}{2008}]{ibar08}
{Ibar} E. et~al., 2008, \mnras, 386, 953

\bibitem[\protect\citeauthoryear{{Ibar} et~al.}{{Ibar} et~al.}{2010}]{ibar10}
{Ibar} E., {Ivison} R.~J., {Best} P.~N., {Coppin} K., {Pope} A., {Smail} I.,
  {Dunlop} J.~S., 2010, \mnras, 401, L53

\bibitem[\protect\citeauthoryear{{Ivison} et~al.}{{Ivison}
  et~al.}{2010a}]{ivison10a}
{Ivison} R.~J. et~al., 2010a, \mnras, 402, 245

\bibitem[\protect\citeauthoryear{{Ivison} et~al.}{{Ivison}
  et~al.}{2010b}]{ivison10b}
{Ivison} R.~J. et~al., 2010b, \aap, 518, L31

\bibitem[\protect\citeauthoryear{{Ivison} et~al.}{{Ivison}
  et~al.}{1998}]{ivison98}
{Ivison} R.~J., {Smail} I., {Le Borgne} J.-F., {Blain} A.~W., {Kneib} J.-P.,
  {Bezecourt} J., {Kerr} T.~H.,  {Davies} J.~K., 1998, \mnras, 298, 583

\bibitem[\protect\citeauthoryear{{Kaiser}}{{Kaiser}}{2006}]{kaiser06}
{Kaiser} C.~R., 2006, \mnras, 367, 1083

\bibitem[\protect\citeauthoryear{{Karim} et~al.}{{Karim}
  et~al.}{2013}]{karim13}
{Karim} A. et~al., 2013, \mnras, 432, 2

\bibitem[\protect\citeauthoryear{{Lacki} \& {Thompson}}{{Lacki} \&
  {Thompson}}{2010}]{lacki10}
{Lacki} B.~C.,  {Thompson} T.~A., 2010, \apj, 717, 196

\bibitem[\protect\citeauthoryear{{Magnelli} et~al.}{{Magnelli}
  et~al.}{2010}]{magnelli10}
{Magnelli} B. et~al., 2010, \aap, 518, L28

\bibitem[\protect\citeauthoryear{{Miller} et~al.}{{Miller}
  et~al.}{2013}]{miller13}
{Miller} N.~A. et~al., 2013, \apjs, 205, 13

\bibitem[\protect\citeauthoryear{{Mullaney} et~al.}{{Mullaney}
  et~al.}{2011}]{mullaney11}
{Mullaney} J.~R., {Alexander} D.~M., {Goulding} A.~D.,  {Hickox} R.~C., 2011,
  \mnras, 414, 1082

\bibitem[\protect\citeauthoryear{{Murphy}}{{Murphy}}{2009}]{murphy09}
{Murphy} E.~J., 2009, \apj, 706, 482

\bibitem[\protect\citeauthoryear{{Murphy}}{{Murphy}}{2013}]{murphy13}
{Murphy} E.~J., 2013, \apj, 777, 58

\bibitem[\protect\citeauthoryear{{O'Dea}}{{O'Dea}}{1998}]{odea98}
{O'Dea} C.~P., 1998, \pasp, 110, 493

\bibitem[\protect\citeauthoryear{{Pilbratt} et~al.}{{Pilbratt}
  et~al.}{2010}]{pilbratt10}
{Pilbratt} G.~L. et~al., 2010, \aap, 518, L1

\bibitem[\protect\citeauthoryear{{Pope} et~al.}{{Pope} et~al.}{2008}]{pope08}
{Pope} A. et~al., 2008, \apj, 675, 1171

\bibitem[\protect\citeauthoryear{{Pope} et~al.}{{Pope} et~al.}{2006}]{pope06}
{Pope} A. et~al., 2006, \mnras, 370, 1185

\bibitem[\protect\citeauthoryear{{Sanders} \& {Mirabel}}{{Sanders} \&
  {Mirabel}}{1996}]{sanders96}
{Sanders} D.~B.,  {Mirabel} I.~F., 1996, \araa, 34, 749

\bibitem[\protect\citeauthoryear{{Sargent} et~al.}{{Sargent}
  et~al.}{2010}]{sargent10}
{Sargent} M.~T. et~al., 2010, \apjl, 714, L190

\bibitem[\protect\citeauthoryear{{Schleicher} \& {Beck}}{{Schleicher} \&
  {Beck}}{2013}]{schleicher13}
{Schleicher} D.~R.~G.,  {Beck} R., 2013, \aap, 556, A142

\bibitem[\protect\citeauthoryear{{Seymour} et~al.}{{Seymour}
  et~al.}{2009}]{seymour09}
{Seymour} N., {Huynh} M., {Dwelly} T., {Symeonidis} M., {Hopkins} A., {McHardy}
  I.~M., {Page} M.~J.,  {Rieke} G., 2009, \mnras, 398, 1573

\bibitem[\protect\citeauthoryear{{Simpson} et~al.}{{Simpson}
  et~al.}{2013}]{simpson13}
{Simpson} J. et~al., 2013, ArXiv e-prints:1310.6363

\bibitem[\protect\citeauthoryear{{Smail} et~al.}{{Smail}
  et~al.}{2000}]{smail00}
{Smail} I., {Ivison} R.~J., {Owen} F.~N., {Blain} A.~W.,  {Kneib} J.-P., 2000,
  \apj, 528, 612

\bibitem[\protect\citeauthoryear{{Swinbank} et~al.}{{Swinbank}
  et~al.}{2014}]{swinbank13}
{Swinbank} A.~M. et~al., 2014, \mnras, 438, 1267

\bibitem[\protect\citeauthoryear{{Valiante} et~al.}{{Valiante}
  et~al.}{2007}]{valiante07}
{Valiante} E., {Lutz} D., {Sturm} E., {Genzel} R., {Tacconi} L.~J., {Lehnert}
  M.~D.,  {Baker} A.~J., 2007, \apj, 660, 1060

\bibitem[\protect\citeauthoryear{{Voelk}}{{Voelk}}{1989}]{voelk89}
{Voelk} H.~J., 1989, \aap, 218, 67

\bibitem[\protect\citeauthoryear{{Wang} et~al.}{{Wang} et~al.}{2013}]{wang13}
{Wang} S.~X. et~al., 2013, \apj, 778, 179

\bibitem[\protect\citeauthoryear{{Wei{\ss}} et~al.}{{Wei{\ss}}
  et~al.}{2009}]{weiss09}
{Wei{\ss}} A. et~al., 2009, \apj, 707, 1201

\bibitem[\protect\citeauthoryear{{Yun}, {Reddy} \& {Condon}}{{Yun}
  et~al.}{2001}]{yun01}
{Yun} M.~S., {Reddy} N.~A.,  {Condon} J.~J., 2001, \apj, 554, 803

\end{thebibliography}

\begin{table*}
\centering
{\small
\caption{The radio properties of selected ALESS SMGs}}
 \begin{tabular}{lcccccccccc}
    \hline
    \multicolumn{1}{l}{ALESS ID}&
    \multicolumn{1}{c}{RA}&
    \multicolumn{1}{c}{Dec}&
    \multicolumn{1}{c}{$z_{\rm phot}^a$}&
    \multicolumn{1}{c}{$S_{ 870\,\mu{\rm m}}^b$}&
    \multicolumn{1}{c}{$S_{\rm 1.4\,GHz}$$^c$}&
    \multicolumn{1}{c}{$S_{\rm 610\,MHz}$$^c$}&
    \multicolumn{1}{c}{$S_{\rm IR}$}&
    \multicolumn{1}{c}{$\alpha^{1400}_{610}$}&
    \multicolumn{1}{c}{q$_{\rm IR}$}&\\
    \multicolumn{1}{l}{}&
    \multicolumn{1}{c}{(J\,2000)}&
    \multicolumn{1}{c}{(J\,2000)}&
    \multicolumn{1}{c}{}&
    \multicolumn{1}{c}{(mJy)}&
    \multicolumn{1}{c}{($\mu$Jy)}&
    \multicolumn{1}{c}{($\mu$Jy)}&
    \multicolumn{1}{c}{($\times 10^{-15}$\,W\,m$^{-2}$)}&
    \multicolumn{1}{c}{}&
    \multicolumn{1}{c}{}&\\
    \hline
    ALESS 001.1 &  53.3103 & $-$             27.9374 & $ 4.34^{+ 2.66}_{- 1.43}$ &  6.7$\pm$                 0.5 & $<$          25 & $<$          45 & $ 0.07\pm 0.16$ & --- & $<$ 2.46 \\ 
ALESS 001.2 &  53.3101 & $-$             27.9366 & $ 4.64^{+ 2.34}_{- 1.02}$ &  3.5$\pm$                 0.4 & $<$          24 & $<$          47 & $ 0.06\pm 0.05$ & --- & $<$ 2.42 \\ 
ALESS 001.3 &  53.3091 & $-$             27.9368 & $ 2.85^{+ 0.20}_{- 0.30}$ &  1.9$\pm$                 0.4 & $<$          23 & $<$          44 & $ 0.18\pm 0.06$ & --- & $<$ 2.79 \\ 
ALESS 002.1 &  53.2612 & $-$             27.9452 & $ 1.96^{+ 0.27}_{- 0.20}$ &  3.8$\pm$                 0.4 &          237$\pm$           8 &          470$\pm$          13 & $ 0.61\pm 0.00$ & $-$                0.82$\pm$                0.03 &  2.23$\pm$ 0.01 \\ 
ALESS 003.1 &  53.3396 & $-$             27.9223 & $ 3.90^{+ 0.50}_{- 0.59}$ &  8.3$\pm$                 0.4 &           29$\pm$           8 &           82$\pm$          17 & $ 0.24\pm 0.12$ & $-$                1.26$\pm$                0.30 &  3.21$\pm$ 0.26 \\ 
ALESS 005.1 &  52.8705 & $-$             27.9858 & $ 2.86^{+ 0.05}_{- 0.04}$ &  7.8$\pm$                 0.7 &           42$\pm$           9 &           91$\pm$          22 & $ 0.27\pm 0.02$ & $-$                0.94$\pm$                0.22 &  2.80$\pm$ 0.10 \\ 
ALESS 006.1 &  53.2373 & $-$             28.0169 & $ 0.45^{+ 0.06}_{- 0.04}$ &  6.0$\pm$                 0.4 &           49$\pm$           8 & $<$          73 & $ 1.04\pm 0.34$ & $>$$-$                0.48 &  2.77$\pm$ 0.16 \\ 
ALESS 007.1 &  53.3142 & $-$             27.7568 & $ 2.50^{+ 0.12}_{- 0.16}$ &  6.1$\pm$                 0.3 &           79$\pm$           7 &          145$\pm$          18 & $ 0.50\pm 0.09$ & $-$                0.74$\pm$                0.09 &  2.63$\pm$ 0.09 \\ 
ALESS 009.1 &  53.0472 & $-$             27.8700 & $ 4.50^{+ 0.54}_{- 2.33}$ &  8.8$\pm$                 0.5 &           35$\pm$           7 &           52$\pm$          16 & $ 0.16\pm 0.13$ & $-$                0.48$\pm$                0.22 &  2.44$\pm$ 0.38 \\ 
ALESS 010.1 &  53.0794 & $-$             27.8708 & $ 2.02^{+ 0.09}_{- 0.09}$ &  5.2$\pm$                 0.5 &           49$\pm$           6 & $<$          71 & $ 0.40\pm 0.02$ & $>$$-$                0.45 &  2.42$\pm$ 0.06 \\ 
ALESS 011.1 &  53.0577 & $-$             27.9334 & $ 2.83^{+ 1.88}_{- 0.50}$ &  7.3$\pm$                 0.4 &           55$\pm$           7 &           98$\pm$          16 & $ 0.45\pm 0.04$ & $-$                0.68$\pm$                0.13 &  2.74$\pm$ 0.07 \\ 
ALESS 013.1 &  53.2041 & $-$             27.7144 & $ 3.25^{+ 0.64}_{- 0.46}$ &  8.0$\pm$                 0.6 &           25$\pm$           6 &           88$\pm$          21 & $ 0.22\pm 0.09$ & $-$                1.52$\pm$                0.26 &  3.33$\pm$ 0.21 \\ 
ALESS 014.1 &  52.9687 & $-$             28.0553 & $ 4.46^{+ 2.54}_{- 0.88}$ &  7.5$\pm$                 0.5 &           91$\pm$           8 &           86$\pm$          17 & $ 0.18\pm 0.01$ &                 0.07$\pm$                0.10 &  1.67$\pm$ 0.05 \\ 
ALESS 015.1 &  53.3890 & $-$             27.9915 & $ 1.93^{+ 0.62}_{- 0.33}$ &  9.0$\pm$                 0.4 &           32$\pm$          10 &           84$\pm$          16 & $ 0.19\pm 0.40$ & $-$                1.16$\pm$                0.31 &  2.73$\pm$ 0.94 \\ 
ALESS 017.1 &  53.0304 & $-$             27.8558 & $ 1.50^{+ 0.10}_{- 0.07}$ &  8.4$\pm$                 0.5 &           90$\pm$           7 &          174$\pm$          28 & $ 0.44\pm 0.12$ & $-$                0.79$\pm$                0.08 &  2.43$\pm$ 0.13 \\ 
ALESS 018.1 &  53.0203 & $-$             27.7799 & $ 2.04^{+ 0.10}_{- 0.06}$ &  4.4$\pm$                 0.5 &           72$\pm$           7 &          291$\pm$          44 & $ 0.68\pm 0.11$ & $-$                1.68$\pm$                0.10 &  3.21$\pm$ 0.08 \\ 
ALESS 019.1 &  53.0344 & $-$             27.9706 & $ 2.41^{+ 0.17}_{- 0.11}$ &  5.0$\pm$                 0.4 &           35$\pm$           7 &           95$\pm$          20 & $ 0.30\pm 0.04$ & $-$                1.21$\pm$                0.20 &  3.01$\pm$ 0.10 \\ 
ALESS 019.2 &  53.0329 & $-$             27.9733 & $ 2.17^{+ 0.09}_{- 0.10}$ &  2.0$\pm$                 0.5 & $<$          26 & $<$          57 & $ 0.16\pm 0.14$ & --- & $<$ 2.62 \\ 
ALESS 022.1 &  52.9455 & $-$             27.5443 & $ 1.88^{+ 0.19}_{- 0.23}$ &  4.5$\pm$                 0.5 &           65$\pm$           9 &          109$\pm$          25 & $ 0.41\pm 0.39$ & $-$                0.62$\pm$                0.14 &  2.51$\pm$ 0.42 \\ 
ALESS 023.1 &  53.0500 & $-$             28.0851 & $ 4.99^{+ 2.01}_{- 2.55}$ &  6.7$\pm$                 0.4 & $<$          32 & $<$          85 & $ 0.10\pm 0.07$ & --- & $<$ 2.53 \\ 
ALESS 025.1 &  52.9870 & $-$             27.9943 & $ 2.23^{+ 0.07}_{- 0.17}$ &  6.2$\pm$                 0.5 &           63$\pm$           8 &           94$\pm$          20 & $ 0.76\pm 0.04$ & $-$                0.47$\pm$                0.13 &  2.74$\pm$ 0.06 \\ 
ALESS 029.1 &  53.4037 & $-$             27.9693 & $ 2.65^{+ 2.94}_{- 0.76}$ &  5.9$\pm$                 0.4 &           50$\pm$          10 &           57$\pm$          18 & $ 0.67\pm 0.06$ & $-$                0.15$\pm$                0.21 &  2.64$\pm$ 0.09 \\ 
ALESS 031.1 &  52.9574 & $-$             27.9613 & $ 2.88^{+ 1.80}_{- 0.41}$ &  8.1$\pm$                 0.4 &           30$\pm$           7 & $<$          65 & $ 0.40\pm 0.29$ & $>$$-$                0.91 &  2.97$\pm$ 0.33 \\ 
ALESS 037.1 &  53.4006 & $-$             27.8974 & $ 3.53^{+ 0.56}_{- 0.31}$ &  2.9$\pm$                 0.4 &           32$\pm$           8 & $<$          50 & $ 0.14\pm 0.01$ & $>$$-$                0.56 &  2.36$\pm$ 0.12 \\ 
ALESS 037.2 &  53.4015 & $-$             27.8967 & $ 4.87^{+ 0.21}_{- 0.40}$ &  1.6$\pm$                 0.4 & $<$          25 & $<$          49 & $ 0.03\pm 0.03$ & --- & $<$ 2.13 \\ 
ALESS 039.1 &  52.9376 & $-$             27.5769 & $ 2.44^{+ 0.18}_{- 0.23}$ &  4.3$\pm$                 0.3 &           50$\pm$           8 &           73$\pm$          23 & $ 0.31\pm 0.04$ & $-$                0.46$\pm$                0.18 &  2.47$\pm$ 0.09 \\ 
ALESS 043.1 &  53.2777 & $-$             27.8007 & $ 1.71^{+ 0.20}_{- 0.12}$ &  2.3$\pm$                 0.4 & $<$          24 & $<$          83 & $ 0.20\pm 0.14$ & --- & $<$ 2.70 \\ 
ALESS 045.1 &  53.1053 & $-$             27.8751 & $ 2.34^{+ 0.26}_{- 0.67}$ &  6.0$\pm$                 0.5 &           36$\pm$           7 & $<$          58 & $ 0.32\pm 0.02$ & $>$$-$                0.59 &  2.66$\pm$ 0.09 \\ 
ALESS 049.1 &  52.8530 & $-$             27.8464 & $ 2.76^{+ 0.11}_{- 0.14}$ &  6.0$\pm$                 0.7 &           85$\pm$           8 &          154$\pm$          50 & $ 0.44\pm 0.05$ & $-$                0.72$\pm$                0.11 &  2.56$\pm$ 0.07 \\ 
ALESS 049.2 &  52.8520 & $-$             27.8439 & $ 1.47^{+ 0.07}_{- 0.10}$ &  1.8$\pm$                 0.5 &           37$\pm$           8 & $<$         138 & $ 0.26\pm 0.21$ & $>$$-$                1.58 &  2.88$\pm$ 0.36 \\ 
ALESS 051.1 &  52.9378 & $-$             27.7409 & $ 1.22^{+ 0.03}_{- 0.06}$ &  4.7$\pm$                 0.4 &           35$\pm$           7 &           78$\pm$          24 & $ 0.38\pm 0.14$ & $-$                0.99$\pm$                0.22 &  2.81$\pm$ 0.18 \\ 
ALESS 055.1 &  53.2592 & $-$             27.6765 & $ 2.05^{+ 0.15}_{- 0.13}$ &  4.0$\pm$                 0.4 & $<$          25 & $<$          52 & $ 0.06\pm 0.00$ & --- & $<$ 2.22 \\ 
ALESS 055.5 &  53.2598 & $-$             27.6765 & $ 2.35^{+ 0.11}_{- 0.13}$ &  1.4$\pm$                 0.4 & $<$          25 & $<$          50 & $ 0.08\pm 0.11$ & --- & $<$ 2.39 \\ 
ALESS 057.1 &  52.9663 & $-$             27.8909 & $ 2.95^{+ 0.05}_{- 0.10}$ &  3.6$\pm$                 0.6 &           51$\pm$           7 &           83$\pm$          23 & $ 0.23\pm 0.04$ & $-$                0.59$\pm$                0.15 &  2.42$\pm$ 0.10 \\ 
ALESS 059.2 &  53.2659 & $-$             27.7384 & $ 2.09^{+ 0.78}_{- 0.29}$ &  1.9$\pm$                 0.4 & $<$          25 & $<$          53 & $ 0.19\pm 0.21$ & --- & $<$ 2.71 \\ 
ALESS 061.1 &  53.1911 & $-$             28.0065 & $ 6.52^{+ 0.36}_{- 0.34}$ &  4.3$\pm$                 0.5 & $<$          26 & $<$          61 & $ 0.05\pm 0.02$ & --- & $<$ 2.38 \\ 
ALESS 063.1 &  53.2852 & $-$             28.0122 & $ 1.86^{+ 0.10}_{- 0.33}$ &  5.6$\pm$                 0.3 & $<$          27 & $<$          58 & $ 0.09\pm 0.01$ & --- & $<$ 2.32 \\ 
ALESS 065.1 &  53.2178 & $-$             27.5906 & $ 2.81^{+ 0.95}_{- 0.36}$ &  4.2$\pm$                 0.4 & $<$          25 & $<$          85 & $ 0.22\pm 0.10$ & --- & $<$ 2.85 \\ 
ALESS 066.1 &  53.3831 & $-$             27.9026 & $ 2.33^{+ 0.05}_{- 0.04}$ &  2.5$\pm$                 0.5 &           70$\pm$           8 & $<$          46 & $ 0.30\pm 0.04$ & $>$                0.51 &  1.64$\pm$ 0.08 \\ 
ALESS 067.1 &  53.1800 & $-$             27.9206 & $ 2.13^{+ 0.05}_{- 0.09}$ &  4.5$\pm$                 0.4 &           74$\pm$           7 &          137$\pm$          15 & $ 0.60\pm 0.08$ & $-$                0.75$\pm$                0.10 &  2.70$\pm$ 0.07 \\ 
ALESS 067.2 &  53.1793 & $-$             27.9207 & $ 2.06^{+ 0.06}_{- 0.16}$ &  1.7$\pm$                 0.4 &           54$\pm$           8 &          137$\pm$          15 & $ 0.14\pm 0.03$ & $-$                1.13$\pm$                0.15 &  2.39$\pm$ 0.12 \\ 
ALESS 069.1 &  52.8907 & $-$             27.9923 & $ 2.34^{+ 0.27}_{- 0.44}$ &  4.9$\pm$                 0.6 &           31$\pm$          10 & $<$          65 & $ 0.20\pm 0.02$ & $>$$-$                0.90 &  2.46$\pm$ 0.15 \\ 
ALESS 070.1 &  52.9334 & $-$             27.6432 & $ 2.27^{+ 0.05}_{- 0.06}$ &  5.2$\pm$                 0.4 &          273$\pm$           8 &          533$\pm$          21 & $ 0.77\pm 0.08$ & $-$                0.81$\pm$                0.03 &  2.29$\pm$ 0.05 \\ 
ALESS 071.1 &  53.2735 & $-$             27.5578 & $ 2.48^{+ 0.21}_{- 0.10}$ &  2.9$\pm$                 0.6 &          199$\pm$           9 &          370$\pm$          21 & $ 1.13\pm 0.00$ & $-$                0.75$\pm$                0.05 &  2.58$\pm$ 0.02 \\ 
ALESS 071.3 &  53.2756 & $-$             27.5564 & $ 2.73^{+ 0.22}_{- 0.25}$ &  1.4$\pm$                 0.4 & $<$          28 & $<$          71 & $ 0.17\pm 0.06$ & --- & $<$ 2.66 \\ 
ALESS 073.1 &  53.1220 & $-$             27.9388 & $ 5.18^{+ 0.43}_{- 0.45}$ &  6.1$\pm$                 0.5 &           24$\pm$           6 & $<$          41 & $ 0.08\pm 0.02$ & $>$$-$                0.64 &  2.32$\pm$ 0.18 \\ 
ALESS 074.1 &  53.2881 & $-$             27.8048 & $ 1.80^{+ 0.13}_{- 0.13}$ &  4.6$\pm$                 0.7 &           48$\pm$           8 & $<$         125 & $ 0.49\pm 0.00$ & $>$$-$                1.15 &  2.20$\pm$ 0.07 \\ 
ALESS 075.1 &  52.8633 & $-$             27.9309 & $ 2.39^{+ 0.08}_{- 0.06}$ &  3.2$\pm$                 0.4 &           75$\pm$           9 &           86$\pm$          24 & $ 0.36\pm 0.08$ & $-$                0.16$\pm$                0.13 &  2.19$\pm$ 0.11 \\ 
ALESS 075.4 &  52.8607 & $-$             27.9321 & $ 2.10^{+ 0.29}_{- 0.34}$ &  1.3$\pm$                 0.4 & $<$          30 & $<$          76 & $ 0.13\pm 0.18$ & --- & $<$ 2.48 \\ 
ALESS 079.1 &  53.0881 & $-$             27.9408 & $ 2.04^{+ 0.62}_{- 0.31}$ &  4.1$\pm$                 0.4 & $<$          25 &           85$\pm$          19 & $ 0.33\pm 0.02$ & $-$                1.46$\pm$                0.34 & $<$ 3.25$\pm$ 0.15 \\ 
ALESS 079.2 &  53.0900 & $-$             27.9400 & $ 1.55^{+ 0.10}_{- 0.18}$ &  2.0$\pm$                 0.4 &           41$\pm$           7 & $<$          59 & $ 0.53\pm 0.16$ & $>$$-$                0.44 &  2.62$\pm$ 0.15 \\ 
ALESS 080.1 &  52.9283 & $-$             27.8102 & $ 1.96^{+ 0.16}_{- 0.14}$ &  4.0$\pm$                 0.9 &           49$\pm$           7 & $<$          78 & $ 0.18\pm 0.05$ & $>$$-$                0.56 &  2.06$\pm$ 0.14 \\ 
ALESS 080.2 &  52.9276 & $-$             27.8114 & $ 1.36^{+ 0.17}_{- 0.08}$ &  3.5$\pm$                 0.9 & $<$          25 & $<$          74 & $ 0.14\pm 0.18$ & --- & $<$ 2.49 \\ 
ALESS 082.1 &  53.2250 & $-$             27.6375 & $ 2.10^{+ 3.27}_{- 0.44}$ &  1.9$\pm$                 0.5 & $<$          24 & $<$          74 & $ 0.22\pm 0.17$ & --- & $<$ 2.77 \\ 
ALESS 083.4 &  53.2863 & $-$             28.0885 & $ 0.56^{+ 1.54}_{- 0.50}$ &  1.4$\pm$                 0.4 & $<$          31 & $<$          60 & $ 0.97\pm 1.71$ & --- & $<$ 3.08 \\ 
ALESS 084.1 &  52.9771 & $-$             27.8516 & $ 1.92^{+ 0.08}_{- 0.07}$ &  3.2$\pm$                 0.6 &           39$\pm$           7 &          122$\pm$          28 & $ 0.40\pm 0.05$ & $-$                1.38$\pm$                0.18 &  3.07$\pm$ 0.09 \\ 
ALESS 084.2 &  52.9744 & $-$             27.8512 & $ 1.75^{+ 0.08}_{- 0.19}$ &  3.2$\pm$                 0.8 & $<$          23 & $<$          79 & $ 0.19\pm 0.08$ & --- & $<$ 2.70 \\ 
ALESS 087.1 &  53.2120 & $-$             27.5282 & $ 3.20^{+ 0.08}_{- 0.47}$ &  1.3$\pm$                 0.4 &           58$\pm$           9 &          171$\pm$          27 & $ 0.14\pm 0.01$ & $-$                1.29$\pm$                0.16 &  2.61$\pm$ 0.08 \\ 
ALESS 088.1 &  52.9782 & $-$             27.8949 & $ 1.83^{+ 0.12}_{- 0.11}$ &  4.6$\pm$                 0.6 &           34$\pm$           7 &          140$\pm$          21 & $ 0.37\pm 0.13$ & $-$                1.72$\pm$                0.20 &  3.24$\pm$ 0.18 \\ 

  \end{tabular}
\end{table*}
\begin{table*}
\centering
{\small
\contcaption{}}
\begin{tabular}{lcccccccccc}
    \hline
    \multicolumn{1}{l}{ALESS ID}&
    \multicolumn{1}{c}{RA}&
    \multicolumn{1}{c}{Dec}&
    \multicolumn{1}{c}{$z_{\rm phot}^a$}&
    \multicolumn{1}{c}{$S_{ 870\,\mu{\rm m}}^b$}&
    \multicolumn{1}{c}{$S_{\rm 1.4\,GHz}$$^c$}&
    \multicolumn{1}{c}{$S_{\rm 610\,MHz}$$^c$}&
    \multicolumn{1}{c}{$S_{\rm IR}$}&
    \multicolumn{1}{c}{$\alpha^{1400}_{610}$}&
    \multicolumn{1}{c}{q$_{\rm IR}$}&\\
    \multicolumn{1}{l}{}&
    \multicolumn{1}{c}{(J\,2000)}&
    \multicolumn{1}{c}{(J\,2000)}&
    \multicolumn{1}{c}{}&
    \multicolumn{1}{c}{(mJy)}&
    \multicolumn{1}{c}{($\mu$Jy)}&
    \multicolumn{1}{c}{($\mu$Jy)}&
    \multicolumn{1}{c}{($\times 10^{-15}$\,W\,m$^{-2}$)}&
    \multicolumn{1}{c}{}&
    \multicolumn{1}{c}{}&\\
    \hline
    ALESS 088.11 &  52.9789 & $-$             27.8938 & $ 2.56^{+ 0.04}_{- 0.12}$ &  2.5$\pm$                 0.7 & $<$          23 & $<$          67 & $ 0.17\pm 0.16$ & --- & $<$ 2.74 \\ 
ALESS 088.5 &  52.9825 & $-$             27.8964 & $ 2.29^{+ 0.11}_{- 0.50}$ &  2.9$\pm$                 0.7 &           38$\pm$           7 & $<$          60 & $ 0.43\pm 0.11$ & $>$$-$                0.55 &  2.68$\pm$ 0.14 \\ 
ALESS 092.2 &  52.9089 & $-$             27.7287 & $ 1.90^{+ 0.27}_{- 0.74}$ &  2.4$\pm$                 0.7 & $<$          24 & $<$          54 & $ 0.06\pm 0.02$ & --- & $<$ 2.21 \\ 
ALESS 094.1 &  53.2816 & $-$             27.9683 & $ 2.87^{+ 0.37}_{- 0.64}$ &  3.2$\pm$                 0.5 &           27$\pm$           9 & $<$          52 & $ 0.10\pm 0.11$ & $>$$-$                0.79 &  0.28$\pm$ 0.47 \\ 
ALESS 098.1 &  52.8747 & $-$             27.9563 & $ 1.63^{+ 0.17}_{- 0.08}$ &  4.8$\pm$                 0.6 &          145$\pm$           8 &          314$\pm$          24 & $ 1.02\pm 0.15$ & $-$                0.93$\pm$                0.06 &  2.67$\pm$ 0.07 \\ 
ALESS 102.1 &  53.3983 & $-$             27.6731 & $ 1.76^{+ 0.16}_{- 0.18}$ &  3.1$\pm$                 0.5 &           39$\pm$           9 &           55$\pm$          14 & $ 0.25\pm 0.14$ & $-$                0.44$\pm$                0.25 &  2.43$\pm$ 0.27 \\ 
ALESS 107.1 &  52.8771 & $-$             27.8636 & $ 3.75^{+ 0.09}_{- 0.08}$ &  1.9$\pm$                 0.4 & $<$          25 & $<$         105 & $ 0.10\pm 0.03$ & --- & $<$ 2.58 \\ 
ALESS 107.3 &  52.8780 & $-$             27.8655 & $ 2.12^{+ 1.54}_{- 0.81}$ &  1.5$\pm$                 0.4 & $<$          25 & $<$          99 & $ 0.13\pm 0.29$ & --- & $<$ 2.52 \\ 
ALESS 110.1 &  52.8444 & $-$             27.9048 & $ 2.54^{+ 0.70}_{- 0.50}$ &  4.1$\pm$                 0.5 &           31$\pm$           9 & $<$          80 & $ 0.39\pm 0.05$ & $>$$-$                1.16 &  2.46$\pm$ 0.14 \\ 
ALESS 114.2 &  52.9629 & $-$             27.7437 & $ 1.55^{+ 0.07}_{- 0.07}$ &  2.0$\pm$                 0.5 &           98$\pm$           7 &          170$\pm$          24 & $ 0.65\pm 0.25$ & $-$                0.66$\pm$                0.08 &  2.52$\pm$ 0.17 \\ 
ALESS 116.1 &  52.9763 & $-$             27.7580 & $ 3.54^{+ 1.47}_{- 0.87}$ &  3.1$\pm$                 0.5 & $<$          23 & $<$         103 & $ 0.09\pm 0.12$ & --- & $<$ 2.53 \\ 
ALESS 116.2 &  52.9768 & $-$             27.7587 & $ 4.01^{+ 1.19}_{- 2.19}$ &  3.4$\pm$                 0.6 &           42$\pm$           7 &          121$\pm$          36 & $ 0.10\pm 0.02$ & $-$                1.28$\pm$                0.17 &  2.71$\pm$ 0.13 \\ 
ALESS 118.1 &  52.8413 & $-$             27.8282 & $ 2.26^{+ 0.50}_{- 0.23}$ &  3.2$\pm$                 0.5 &           44$\pm$           8 & $<$          81 & $ 0.23\pm 0.07$ & $>$$-$                0.74 &  1.58$\pm$ 0.16 \\ 
ALESS 119.1 &  53.2360 & $-$             28.0570 & $ 3.50^{+ 0.95}_{- 0.35}$ &  8.3$\pm$                 0.5 &           38$\pm$           9 &           74$\pm$          21 & $ 0.15\pm 0.13$ & $-$                0.80$\pm$                0.24 &  2.53$\pm$ 0.39 \\ 
ALESS 122.1 &  52.9148 & $-$             27.6888 & $ 2.06^{+ 0.05}_{- 0.06}$ &  3.7$\pm$                 0.4 &          172$\pm$           7 &          341$\pm$          17 & $ 0.78\pm 0.06$ & $-$                0.83$\pm$                0.04 &  2.48$\pm$ 0.04 \\ 
ALESS 124.1 &  53.0168 & $-$             27.6018 & $ 6.07^{+ 0.93}_{- 1.16}$ &  3.6$\pm$                 0.6 &           28$\pm$           8 &           99$\pm$          24 & $ 0.07\pm 0.01$ & $-$                1.54$\pm$                0.29 &  3.12$\pm$ 0.14 \\ 
ALESS 126.1 &  53.0400 & $-$             27.6855 & $ 1.81^{+ 0.28}_{- 0.08}$ &  2.2$\pm$                 0.5 &           25$\pm$           7 & $<$          80 & $ 0.22\pm 0.08$ & $>$$-$                1.40 &  2.88$\pm$ 0.19 \\ 

    \hline
  \end{tabular}
{\small
\\Description of column headings: ALESS ID; RA; Dec; photometric redshift; ALMA 870\,$\mu$m; VLA 1.4\,GHz; and GMRT 610\,MHz flux densities; infrared flux ($S_{\rm IR}=L_{\rm IR}/4\pi R^2$); 1.4\,GHz-to-610\,MHz spectral index, $\alpha$; and infrared/radio flux ratio, $q_{\rm IR}$ (see text for definition). Notes\: $^a$From \citet{simpson13}; $^b$ From \citet{swinbank13};$^c$ Non-detections are represented by $3\sigma$ flux limits.}
\end{table*}

\label{lastpage}
\enddocument